\documentclass[reprint,nofootinbib,aps,prd,showpacs,superscriptaddress,groupedaddress,amsmath,longbibliography]{revtex4-1}
\usepackage{braket}
\usepackage{graphicx}
\usepackage{dcolumn}   
\usepackage{bm}
\usepackage{url}
\usepackage{multirow}
\usepackage{footnote}
\usepackage{epstopdf}
\usepackage{hyperref}
\usepackage{xcolor}

\definecolor{darkblue}{cmyk}{1, 1, 0, 0}
\hypersetup{colorlinks=true,urlcolor=darkblue,citecolor=darkblue,linkcolor=darkblue}

\newcommand{\physrep}{Phys.~Rep.}
\newcommand{\aap}{Astronomy \& Astrophysics}
\newcommand{\jcap}{Journal of Cosmology and Astroparticle Physics}
\newcommand{\mnras}{Mon.~Not.~R.~Astron.~Soc.}

\newcolumntype{C}[1]{>{\centering\let\newline\\\arraybackslash\hspace{0pt}}m{#1}}
\newcolumntype{L}[1]{>{\raggedright\let\newline\\\arraybackslash\hspace{0pt}}m{#1}}
\newcolumntype{R}[1]{>{\raggedleft\let\newline\\\arraybackslash\hspace{0pt}}m{#1}}
\newcommand{\AddPhantomMinusIfNeeded}[1]{%
\IfDecimal{#1}{
    \IfBeginWith{#1}{-}{\ensuremath{#1}}{\ensuremath{\phantom{-}#1}}%
    }{%
        \ensuremath{\phantom{-}}#1
    }%
}%

\begin{document}

\title{Testing gravity theories using tensor perturbations}
\date{\today}
\author{Weikang Lin}
\email{wxl123830@utdallas.edu}
\author{Mustapha Ishak}
\email{mishak@utdallas.edu}
\affiliation{Department of Physics, The University of Texas at Dallas, Richardson, Texas 75083, USA}

\begin{abstract}
Primordial gravitational waves constitute a promising probe of the very early Universe and the laws of gravity. We study in this work changes to tensor-mode perturbations that can arise in various proposed modified gravity theories. These include additional friction effects, nonstandard dispersion relations involving a massive graviton, a modified speed, and a small-scale modification. We introduce a physically motivated parametrization of these effects and use current available data to obtain exclusion regions in the parameter spaces. Taking into account the foreground subtraction, we then perform a forecast analysis focusing on the tensor-mode modified-gravity parameters as constrained by future experiments COrE, Stage-IV and PIXIE.  For a fiducial value of the tensor-to-scalar ratio $r=0.01$, we find that an additional friction of $3.5\sim4.5\%$ compared to GR will be detected at $3$-$\sigma$ by these experiments, while a decrease in friction will be more difficult to detect. The speed of gravitational waves needs to be by $5\sim15\%$ different from the speed of light for detection. We find that the minimum detectable graviton mass is about $7.8\sim9.7\times 10^{-33}\,eV$, which is of the same order of magnitude as the graviton mass that allows massive gravity theories to produce late-time cosmic acceleration. Finally, we study the tensor-mode perturbations in modified gravity during inflation using our parametrization. We find that, in addition to being related to $r$, the tensor spectral index would be related to the friction parameter $\nu_0$ by $n_T=-3\nu_0-r/8$. Assuming that the friction parameter is unchanged throughout the history of the Universe and that it is much larger than $r$, the future experiments considered here will be able to distinguish this modified-gravity consistency relation from the standard inflation consistency relation, and thus can be used as a further test of modified gravity. In summary, tensor-mode perturbations and cosmic-microwave-background B-mode polarization provide a complementary avenue to test gravity theories.
\end{abstract}


\maketitle

\section{Introduction}
Current problems in cosmology such as cosmic acceleration, or older motivations such as finding unified theories of physics have led to searches and proposals of theories of gravity beyond General Relativity (GR). Associated with these proposals are efforts to test GR using cosmological probes. See, for example  \cite{M.G.and.cosmology,Cosmo.test.of.GR,2015-rev-Joyce-Jain,review-PPF-Joyce2016,2015-rev-Berti-etal,review-PPF-Baker2013,2016-rev-Dodelson} for reviews on testing modifications to gravity at cosmological scales. In doing so, instead of building frameworks to test individual modified gravity models, a common and reasonable approach is to parametrize and test departures from general relativity predictions. This approach is well justified in view of the success of the relativistic $\Lambda$ cold dark matter ($\Lambda$CDM) standard model when compared to observations so that any deviation from GR should be small. It can be viewed as simply testing GR with no reference to any modified gravity models. Any difference in the model parameters from their standard values in GR can point us to the right direction of modification to GR.
One could also argue that an efficient parametrization should meet some minimum criteria. First, it should obviously reduce to GR in some limit or given point. Second, it should assemble the behaviors of more than one theory of modified gravity. Third, the parametrization should be minimum so that the possibly captured deviation is not merely due to the increased degrees of freedom to fit the data. And finally the parametrization should allow us to easily assign physical meanings to the parameters.

There has been a considerable amount of work to systematically parametrize scalar-mode-perturbation deviations from GR in the literature, and we refer readers to some reviews on the topic \cite{M.G.and.cosmology,Cosmo.test.of.GR,2015-rev-Joyce-Jain,review-PPF-Joyce2016,2015-rev-Berti-etal,review-PPF-Baker2013,2016-rev-Dodelson, review-PPF-Hu2007} and publicly available codes to perform such tests \cite{MG-camb-cosmomc,ISiTGR-cite1}. On the other hand, the tensor-mode parametrization for modified gravity has not been systematically nor extensively studied, although several non-GR behaviors in the tensor sector have been individually investigated  \cite{Anisotropy.and.MGW,friction,MassiveGinCMB,MG.Ct.of.GW,MGW.on.Bmode}. It is worth mentioning that methods of parametrization come also with some limitations \cite{2015-Linderetal-EFT,2013-Dossett-Ishak-DEP}, nevertheless they can be informative in some cases.

In this paper, we aim to provide a systematic study of tensor-mode modified-gravity (MG) parameters including current bounds on the parameters and future constraints. In Sec. \ref{section-parametrization}, we discuss a general form of the modified tensor-mode propagation equation including different physical effects. In Sec. \ref{section-inflation}, we investigate the tensor-mode perturbations during inflation for two of our parametrization schemes. In Sec. \ref{section-effect}, we illustrate the effects of our MG parameters on the cosmic-microwave-background (CMB) B-mode polarization. In Sec. \ref{section-constraints} we use the available BKP  \cite{BKP2014} and Planck 2015 \cite{Planck2015XIII-Cos.Param.} data to put bounds on the parameter spaces. In Sec. \ref{section-forecast}, we analyze and provide a forecast of constraints on our tensor-mode MG parameters from some future experiments. Finally, we summarize in Sec. \ref{section-summary}.

\section{Tensor Modes in Modified Gravity and their Parametrization}\label{section-parametrization}
Scalar-, vector- and tensor-mode perturbations with respect to rotation symmetry can be treated separately \cite{S.W.cosmo.,S.D.modern.cosmo.}. The line element only with tensor-mode perturbations reads,
\begin{equation}\label{eq-metric-tensor}
ds^2=-dt^2+a^2(t)(\delta_{ij}+D_{ij}(\mathbf{x},t))dx^idx^j ~,
\end{equation}
where $D_{ij}$ is the traceless (i.e., $D_{ii}=0$) and transverse (or divergenceless, i.e., $\partial_iD_{ij}=0$) part of the perturbed metric, $t$ is the cosmic time (or the comoving time), and $a(t)$ is the scale factor. When working in Fourier space, the propagation equation for a mode with a comoving wave number $ k$ and with either helicity ($\lambda=\pm2$) takes the following form, 
\begin{equation}\label{eq-tensoreqGR}
\ddot{h}_{k}+3\frac{\dot{a}}{a}\dot{h}_{k}+\frac{k^2}{a^2}h_{k}=16\pi G\Pi_{k}^T~,
\end{equation}
where $\dot{h}\equiv\frac{dh}{dt}$, and $\Pi_{k}^T$ is the tensor part (i.e., traceless and divergenceless) of the perturbed energy-stress tensor in Fourier space. Since the above equation does not depend on the helicity $\lambda$, we have dropped it from the subscript, but we still keep the subscript $ k$ to remind us that the amplitude is a function of the wavenmuber. We can see from Eq. \eqref{eq-tensoreqGR} that the dynamics of the tensor-mode amplitude for each mode behaves like a damping harmonic oscillator with a source. The second term $3\tfrac{\dot{a}}{a}\dot{h}_{k}$ represents the damping effect (or the friction) caused by the cosmic expansion. The third term $\frac{k^2}{a^2}h_{k}$ means that the frequency of a free wave $\omega_T$ is the same as its physical wave number $\tfrac{k}{a}$, which consequently means that gravitational waves propagate at the speed of light. The term on the right-hand side represents the source that comes from the tensor part of the stress-energy anisotropy. In GR, the effects from the source on the dynamics of the tensor-mode perturbations are small \cite[chapter 6.6]{S.W.cosmo.}, and we assume this is also true in MG. So we ignore the source term and assume the major modification to the tensor-mode perturbations is from the change to the free propagation equation, i.e., the left-hand side of Eq. \eqref{eq-tensoreqGR}.  Here a test particle is assumed to follow a geodesic as in GR and there will be no modification to the Boltzmann equations.

Relativistic theories of gravity other than GR can (\textit{i}) change the damping rate of gravitational waves (i.e., the term with $\dot{h}$ in the propagation equation), (\textit{ii})  modify the dispersion relation (i.e., rather than $ k^2/a^2$ in the third term, it can be a generic function of $ k/a$; see for example the Ho$\check{\mathrm{r}}$ava-Lifshitz gravity  \cite{tensor.in.HoravaLifshitz} and the Einstein-{\ae}ther theory  \cite{ein.aether.waves}), and (\textit{iii}) add an additional source term on the right-hand side even in the situation of a perfect fluid (see, for example, in the generalized single scalar field theory  \cite{Pertur.in.Ginflation,pertur.in.scalarfield}, and a recent extension to the Horndeski theories \cite{GLPV,GLPV2,GLPV.perturbation}).  Ignoring the source term as we assume it gives small effects, we suggest in this paper the following practical form of the modified propagation equation for tensor-mode perturbations,
\begin{equation}\label{eq-Mo.propagation}
\ddot{h}_{k}+3\frac{\dot{g}}{g}\dot{h}_{k}+\omega_T^2h_{k}=0~,
\end{equation}
where $g$ is a model-dependent function of time via some background variables and is $ k$ independent in the linear regime, and $\omega_T^2$ depends on time and the physical wave number $ k/a$. Similar modified equations are found in the literature  \cite{Anisotropy.and.MGW,friction,MG.Ct.of.GW,MGW.on.Bmode}. In particular, in some previous papers the coefficient in the $\dot{h}$ term has been modified to $(3+\alpha_M)H$ instead of $3H$, which corresponds to $g=a^{1+\frac{\alpha_M}{3}}$ with a constant $\alpha_M$ in Eq. \eqref{eq-Mo.propagation}. For the dispersion relation, a modified speed and a graviton mass have also been considered in the literature.  But here we introduce and use a specific form [Eq. \eqref{eq-Mo.propagation}] based on a more generic friction term and modified dispersion relation. A different parametrization scheme is considered in Ref. \cite{laji.paper}, in which the friction term and the source term are modified in a way that they are both time and waven-umber dependent. This is different from our consideration: 1. We argue that the friction term is only time dependent via some background variables. 2. We neglect changes to the source term since we assume that the effect due to those changes is small in MG. 3. We consider a more general dispersion relation.

Our proposed form of the friction term has more analytical advantages, because it can represent the general friction term for a wide range of MG theories. For example, in $f(R)$ theories (with $R$ being the Ricci scalar), $g=\sqrt{f_R}\times a$, where $f_R=\frac{df(R)}{dR}$ and equals $1$ in GR. In the Horndeski models, we can combine Eq. (5) and Eq. (6) in Ref. \cite{MGW.on.Bmode} and manipulate to get $g=\omega_1^{1/3}\times a$. In tensor-vector-scalar theory, we can modify Eq. (163) in Ref. \cite{TeVeS-cosmology} and get $g=b\gamma$. For all MG theories, the function $g$ depends only on time but not on the wave number.

Our consideration of the modified dispersion relation can in principle cover more generic cases, and is not limited to a constant modified speed $c_T$ or a graviton mass $\mu$. The proposed form of the dispersion relation in Ref. \cite{Anisotropy.and.MGW} reads,
\begin{equation}
\omega_T^2=c_T^2\frac{k^2}{a^2}+\mu^2 ,~\label{eq-dispersion-a}
\end{equation}
which can be manipulated and written as,
\begin{equation}
\frac{\omega_T^2}{k^2/a^2}-1=(c_T^2-1)+\frac{a^2}{k^2}\mu^2~. \label{eq-dispersion-b}
\end{equation}
Here we can see clearly from Eq. \eqref{eq-dispersion-a} or Eq. \eqref{eq-dispersion-b} that the difference from a standard dispersion (i.e., $\frac{\omega_T^2}{k^2/a^2}-1=0$) can be caused by a modified speed $c_T\neq1$ or by a nonzero mass $\mu\neq0$. Note that the squared phase speed of gravitational waves is actually $\frac{\omega_T^2}{(k/a)^2}$, which is different from the squared speed $c_T^2$. In this work, we parametrize the dispersion relation from a different approach. Our starting point of the dispersion-relation parametrization is to treat the right-hand side of Eq. \eqref{eq-dispersion-b} as a whole and small piece.  But we will see that, under a few assumptions, our parametrized dispersion relation corresponds to three physical cases: a modified speed, a graviton mass, and (in addition) an ultraviolet (high-$ k/a$ or small-scale) modification.

There are already some constraints on the dispersion relation in the literature. First, the consideration of gravitational Cherenkov radiation puts a strong lower limit on the phase speed of gravitational waves, which is very close to the speed of light \cite{GW.Cherenkov}. The idea is that, if the phase speed is slower than the speed of light, there must be some energetic particles moving faster than the phase speed of gravitational waves which leads to gravitational Cherenkov radiation. Such gravitational Cherenkov radiation should in principle slow down these energetic particles. But the observed energetic particles can have a speed close to the speed of light, and do not appear to have been slowed down by this process. Or, such particles can only have traveled for a short distance, which contradicts the assumption that they are from the Galactic center or other further sources. In other words, if the idea of gravitational Cherenkov radiation is correct, a subluminal phase speed of gravitational waves is not allowed. Second, for the graviton mass, Ref. \cite{MassiveGinCMB} estimated an upper limit from the CMB observations for a nonvanishing tensor-to-scalar ratio. This bound of graviton mass is stronger than those set by the gravitational-wave detectors. For a more comprehensive list of observational bounds of the graviton mass, we refer readers to Ref. \cite{2016deRham-et-al-graviton-mass-bounds}. 
In this work, however, we will release the above constraints on the dispersion relation. We do so in order to give independent constraints on the tensor sector solely from a Monte Carlo Markov Chain (MCMC) analysis on the current CMB observations.

Now we turn to our parametrization. We first parametrize the dispersion relation. Instead of starting with modifying the speed and adding a graviton mass, we parametrize the dispersion relation from a mathematical point of view. We assume that the dispersion relation depends only on the physical wave number $ k/a$, but not explicitly on time. A general modified dispersion relation that only depends on the physical wave number $ k/a$ takes the following form:
\begin{equation}\label{eq-dispersion1}
\frac{\omega_T^2}{k^2/a^2}-1=\varepsilon(k/a)~,
\end{equation}
where $\varepsilon(k/a)$ is an arbitrary function of $ k/a$ which vanishes in GR. In the last step, we have denoted everything on the right-hand side of \eqref{eq-dispersion-b} as $\varepsilon(k/a)$. This arrangement is motivated by the fact that the deviation from GR is small in the scalar sector, and so we assume the deviation is also small in the tensor sector. A positive/negative $\varepsilon$ corresponds to a superluminal/subluminal phase speed. To parametrize the $ k/a$ dependence of the dispersion relation, we model it such that the deviation either happens in the large-scale or the small-scale limit but unchanged on the other limit, or the deviation is $ k/a$ independent. And the dispersion relation should be isotropic, so it should be an even function of $ k/a$. Under the above assumptions, the following proposals can capture the deviation up to the lowest order, (and there are examples of theories corresponding to each of the following cases,)
\begin{equation}\label{eq-dispersion-cases}
\varepsilon(k/a)=
\begin{cases}
\varepsilon_h\left(\frac{k/a}{K_0}\right)^2~, & \text{small scales,}\\
\varepsilon_0~, & \text{$k/a$ independent,}\\
(\varepsilon_l)^n\left(\frac{\mu_0}{k/a}\right)^2~, & \text{large scales.}
\end{cases}
\end{equation}
In the above, $\varepsilon_0$, $\varepsilon_h$ and $\varepsilon_l$ are tensor-mode MG parameters. The subscripts $h$ and $l$ stand for high- and low- physical wave numbers representively. $ K_0$ and $\mu_0$ are normalization constants. They are inserted to make $\varepsilon_h$ and $\varepsilon_l$ dimensionless and within a practical range (i.e., of unity). For consistency of the units, $k$ in \textsc{camb} is measured in Mpc$^{-1}$, so $ K_0$ and $\mu_0$ is also in Mpc$^{-1}$. There are examples of modified gravity theories that have a dispersion relation in each of the three forms in Eq. \eqref{eq-dispersion-cases}. The first case is a ultraviolet deviation. For example in the Ho\v{r}ava-Lifshitz theory, the dispersion relation deviates from the standard one at small scales  \cite{tensor.in.HoravaLifshitz}, which falls into the first case to the leading order. More explicitly, in Ref. \cite{tensor.in.HoravaLifshitz}, $\frac{K_0}{\varepsilon_h^2}=\frac{g_3}{\zeta^2}$ to the leading order at moderately small scales. The second case corresponds to a constant nonstandard speed of gravitational waves, which can be found in the Einstein-{\ae}ther theory \cite{ein.aether.waves,Anisotropy.and.MGW}. For the third case, an example of deviation happening at large scales is when a graviton mass is added to the propagation equation, $\omega_T^2=\frac{k^2}{a^2}+\mu^2$, which can be written as $\frac{\omega_T^2}{k^2/a^2}-1=\frac{\mu^2}{k^2/a^2}$. And we can identify $(\varepsilon_l)^n$ as the ratio $\mu^2/\mu_0^2$ in the last case. Then our modified dispersion relation is divided into three separate cases, each of which has one parameter, namely $\varepsilon_0$, $\varepsilon_l$ and $\varepsilon_h$. The three parameters characterizing the modified dispersion relation vanish in GR.

\begin{table*}[htpb!]
\begin{ruledtabular}
\caption{\label{table-all-parameters}Table of the tensor-mode MG parameters and their corresponding physical meanings or typical examples. In this work, we consider the four MG parameters separately. Each MG parameter corresponds to a one-parameter modification. All parameters vanish in GR. The physical ranges will be discussed in Sec. \ref{section-effect}.}
\begin{tabular}{lcllc}
Parameters & Scales of deviation & \multicolumn{1}{c}{Physical Meaning or example} & Physical ranges & GR values\\
\hline
$\nu_0$ & All scales & Modulating the friction & $~~~~>-1$ & \multirow{4}{*}{0}\\
$\varepsilon_h$ & Small scales & High $\frac{k}{a}$ deviation, like in Ref. \cite{tensor.in.HoravaLifshitz}& $~~~~~~\geq0$ &\\
$\varepsilon_0$ & All scales & Gives a modified speed & $~~~~~~>-1$ &\\
$\varepsilon_l$ & Large scales & Gives a finite graviton mass & $~~~~~~\geq0$ & \\
\end{tabular}
\end{ruledtabular}
\end{table*}

For the first case, we find $ K_0=100$ Mpc$^{-1}$ suitable. Roughly speaking, $ K_0/\sqrt{\varepsilon_h}$ is the physical wave number onset of the small-scale deviation. In the last case we use $(\varepsilon_l)^n$ instead of simply $\varepsilon_l$, and we set $n=4$. That is because the current constraint on the graviton mass is very weak (to be explored in Sec. \ref{section-constraints}), and it can span four orders of   magnitude. Using $(\varepsilon_l)^4$ roughly make different orders of magnitude of $\varepsilon_l$ at the same footing when using \textsc{CosmoMC}. If further data can provide stronger constraints, we can set $n$ to be a smaller value, for example $n=1$. A value of $\mu_0=1$ Mpc$^{-1}$ corresponds to a graviton mass of $\sim5\times10^{-58}M_p$ in the Planck units, or $\sim6\times10^{-30}\,eV$. In Ref. \cite{MassiveGinCMB}, they used $3000H_0$ (the expansion rate at recombination), which is roughly $0.7$ Mpc$^{-1}$ and this suggests $\mu_0=1$ Mpc$^{-1}$ is suitable.  Any other choices of $ K_0$ and $\mu_0$ can be absorbed into the constants $\varepsilon_h$ and $\varepsilon_l$.

The necessity of the case separation in eq \eqref{eq-dispersion-cases} needs to be justified. We concede that separating the dispersion relation into cases increases the complexity of the analysis. It might not be useful if we only have data corresponding a narrow range of $k/a$, because we would not be able to determine any dependence on $k/a$ from the data. And such case separation does not represent a more general situation where the deviation can occur at both small and large scales. However, the above separation clearly describes different physics of the possible deviations, making it possible to quickly link the modified parameters and the reason for their nonvanishing values. Also for a practical reason, the constraints on the tensor sector are very weak, so it is unrealistic to consider the three deviations simultaneously. One might want to replace the three cases with a power index, such as $(k/a)^n$. Then the positive, zero and negative values of $n$ can generalize the above three cases. But a continuous $n$ lacks physical meaning and can lead to confusion. Therefore, we choose to separate the dispersion relation into three cases.

For the friction term, we simply assume $g=a^{1+\nu_0}$ for a constant $\nu_0$, which is equivalent to the work in Ref. \cite{Anisotropy.and.MGW,friction} as explained earlier in this section. A positive/negative $\nu_0$ means the friction is larger/smaller than the one in GR, and consequently the gravitational waves are more/less damped.

In summary, the MG parameters $\nu_0$, $\varepsilon_0$, $\varepsilon_l$ and $\varepsilon_h$ characterize the modified gravitational-wave-propagation equation in four different cases, and they all vanish in GR. When considered separately (as in this work), the four MG parameters correspond to four one-parameter modifications. The tensor-mode MG parameters and the corresponding physical meanings are summarized in Table \ref{table-all-parameters}.

\section{tensor-mode perturbations during inflation with constant friction and speed}\label{section-inflation}
Our parametrization of the friction term has more analytical advantages. One example is the study of tensor-mode perturbations during inflation. For the case with only a constant friction parameter $\nu_0$, Eq. \eqref{eq-Mo.propagation} in conformal time $d\tau=dt/a$ reads,
\begin{equation}\label{eq-infla1}
h_k''+2\frac{\tilde{g}'}{\tilde{g}}h_k'+k^2h_k=0~,
\end{equation}
where $\tilde{g}=a^{(1+\tilde{\nu_0})}$ for a constant $\tilde{\nu_0}$ and $'$ stands for derivative with respect to the conformal time. Note that, the constant $\tilde{\nu_0}$ in Eq. \eqref{eq-Mo.propagation} is different from the one in Eq. \eqref{eq-infla1}. But they are simply related to each other, and $\tilde{\nu_0}=\frac{3}{2}\nu_0$.  When we let $W = \tilde{g}\times h_k$, Eq. \eqref{eq-infla1} takes the canonical form,
\begin{equation}\label{eq-infla2}
W''+(k^2-\frac{\tilde{g}''}{\tilde{g}})W=0.
\end{equation}
At the early time of inflation when perturbations were inside the horizon, Eq. \eqref{eq-infla2} and $W=\tilde{g}\times h_k$ suggest that the solution is normalized such that,
\begin{equation}\label{eq-normalization}
h_k(t)\rightarrow\frac{\sqrt{16\pi G}}{(2\pi)^{3/2}\sqrt{2k}\tilde{g}}\exp(-ik\int d\tau)~.
\end{equation}
The difference from GR is that we have $\tilde{g}$ in the denominator instead of the scale factor $a$. We assume the Universe was in the ground state so that Eq. \eqref{eq-normalization} will serve as an asymptotic initial condition of $h_k$. To get $h_k$ outside the horizon (by the end of inflation), we need to know the expansion background. Here we first assume the background is exactly exponentially expanding with respect to the cosmic time $t$ (i.e., de Sitter background). We make this assumption at first in order to isolate the MG effects from the slow-roll inflation. Under this assumption, we have $a=-\frac{1}{H\tau}$, where $H$ is the constant expansion rate during inflation. And Eq. \eqref{eq-infla1} becomes,
\begin{equation}\label{eq-infla3}
h_k''-\frac{2(1+\tilde{\nu_0})}{\tau}h_k'+k^2h_k=0.
\end{equation}
If we let $x=-k\tau$ and $h_k=x^{\frac{3}{2}+\tilde{\nu_0}}y$, the above equation becomes,
\begin{equation}\label{eq-bessel}
x^2\frac{d^2y}{dx^2}+x\frac{dy}{dx}+[x^2-(\frac{3}{2}+\tilde{\nu_0})^2]y=0~,
\end{equation}
which is a Bessel differential equation of order $\nu=\frac{3}{2}+\tilde{\nu_0}$ (and this is the reason we use the notation $\nu_0$). The general solution of \eqref{eq-bessel} is a linear combination of Hankel functions of the first and second kinds $H_\nu^{(1)}$ and $H_\nu^{(2)}$. Matching the solution deep inside the horizon [Eq. \eqref{eq-normalization}], we eliminate the $H_\nu^{(2)}$ component since $H_\nu^{(1)}(-k\tau)$ already goes as $\sim\exp(-ik\tau)$. And by taking the outside horizon limit $-k\tau\rightarrow\infty$, we obtain the tensor-mode spectrum,
\begin{equation}\label{eq-tensor-spectrum}
|h_k^0|^2 = \frac{G(2H)^{2(1+\tilde{\nu_0})}\left[\Gamma(\frac{3}{2}+\tilde{\nu_0})\right]^2}{\pi^3\cdot k^{3+2\tilde{\nu_0}}}~.
\end{equation}
where $G$ is the Newtonian constant. The result in GR in a de Sitter background is recovered for $\tilde{\nu_0}=0$. Since $|h_k^0|^2$ is proportional to $k^{-3-2\tilde{\nu_0}}$, we can identify the tensor spectral index as,
\begin{equation}\label{eq-friction-tensor-index}
n_T=-2\tilde{\nu_0}=-3\nu_0~.
\end{equation}
So if $a\propto e^{Ht}$ during inflation, $n_T$ and $\nu_0$ should be related by \eqref{eq-friction-tensor-index}.

For the case of slow-roll inflation, the background is not exactly de Sitter and $H$ is not a constant. One of the slow-roll parameters $\epsilon$ (not one of our modified gravity parameters) measures the first derivative of $H$ with respect to time,
\begin{equation}\label{eq-slow-roll-param}
\epsilon=-\dot{H}/H^2~.
\end{equation}
In this case, the scale factor $a$ no longer goes as $a=-\frac{1}{H\tau}$. Instead it is replaced by $aH=-\frac{1}{(1-\epsilon)\tau}$, which is obtained by integrating Eq. \eqref{eq-slow-roll-param}. As a result, Eq. \eqref{eq-infla3} becomes,
\begin{equation}\label{eq-infla-slow-MG}
h_k''-\frac{2(1+\tilde{\nu_0})}{(1-\epsilon)\tau}h_k'+k^2h_k=0.
\end{equation}
For a small $\epsilon$, we have $\frac{1}{1-\epsilon}\simeq1+\epsilon$, and Eq. \eqref{eq-infla-slow-MG} can be approximately written as,
\begin{equation}\label{eq-slow-roll-param2}
h_k''-\frac{2(1+\tilde{\nu_0}+\epsilon)}{\tau}h_k'+k^2h_k=0.
\end{equation}
Note that $\tilde{\nu_0}$ in \eqref{eq-infla3} is now replaced by $\tilde{\nu_0}+\epsilon$ in \eqref{eq-slow-roll-param2}. Consequently, we only need to replace $\tilde{\nu_0}$ by $\tilde{\nu_0}+\epsilon$ in the final result, i.e., in Eq. \eqref{eq-tensor-spectrum}. In particular, the tensor spectrum index $n_T$ is related to both the MG friction parameter $\nu_0=\tfrac{2}{3}\tilde{\nu_0}$ and the slow-roll parameter $\epsilon$ by,
\begin{equation}\label{eq-MG-nT-fric-slow}
  n_T=-3\nu_0-2\epsilon~.
\end{equation}
In contrast, the ordinary slow-roll inflation in GR gives $n_T=-2\epsilon$ \cite{S.W.cosmo.}. We can see from \eqref{eq-MG-nT-fric-slow} that the MG friction parameter $\nu_0$ and the slow-roll parameter $\epsilon$ have degenerate roles in the tensor spectral index $n_T$. This means the value of $n_T$ can not tell us whether the background is exactly de Sitter with an MG friction parameter $\nu_0$, or slowly changing with a small slow-roll parameter $\epsilon$. The slow-roll inflation consistency relation,
\begin{equation}\label{eq-slow-roll-consistency-rel}
n_T=-r/8~,
\end{equation}
 is expected to change if the friction parameter $\nu_0$ is nonzero. More explicitly, if we assume the result of the scalar sector is unchanged, the tensor-to-scalar ratio $r$ is still related to the slow-roll parameter $\epsilon$ by,
\begin{equation}\label{eq-r-epsilon}
  r=16\epsilon~.
\end{equation}
Note that we have used the fact that the tensor-mode amplitude is not affected by $\nu_0$ to the leading order. Then the inflation consistency relation is now modified in MG and becomes,
\begin{equation}\label{eq-MG-consistency-rel}
  n_T=-3\nu_0-r/8.
\end{equation}
We call Eq. \eqref{eq-MG-consistency-rel} the modified-gravity inflation consistency relation (MG consistency relation).

Verifying the inflation consistency relation is one of the important tasks for future CMB experiments. However the near-future experiments have limited capability of doing so \cite{COrEWhitePaper,PIXIE2011,CMB-Bmode-forecast2015}. The presence of $\nu_0$ in the MG consistency relation \eqref{eq-MG-consistency-rel} makes the situation even worse. For example, if future experiments falsify the standard consistency relation $n_T=-r/8$, it does not necessarily mean the slow-roll inflation is wrong: it can be that general relativity needs to be modified so that the friction term is changed.

It will be difficult for the near-future CMB experiments to disentangle the standard and the MG consistency relations. However, in some extreme cases, the two consistency relations are very different, and this will help us to tell which consistency relation is possibly correct. We explain as follows. The current upper bound of the tensor-to-scalar ratio $r$ is around $0.1$ \cite{BKP2014}. If the true value of $\nu_0$ is much larger than $r$, we can ignore the term $-r/8$ in the MG consistency relation \eqref{eq-MG-consistency-rel}. Then the tensor spectral index reduces to $n_T\simeq-3\nu_0$ in modified gravity. In contrast, the standard consistency relation still gives $n_T=-r/8$. In this case, the MG consistency relation expects $n_T$ to be much larger than what is expected in GR. In the future, if we see $n_T\simeq-3\nu_0$ with $\nu_0\gg r$, then we can say the MG consistency relation is possibly right (or the slow-roll inflation theory has some troubles). In Sec. \ref{subsection-forecast-cons-rel}, we explore how future experiments can distinguish the standard and the MG consistency relations. For the forecast in Sec. \ref{subsection-forecast-cons-rel}, we set for our fiducial model $r=0.01$ and $\nu_0=0.2$. We can then ignore the term $-r/8$ in the MG consistency relation, so $n_T=-3\nu_0-r/8\simeq-3\nu_0 = -0.6$. In contrast, the standard consistency relation in GR is $n_T = -r/8 = -0.00125$. So the values of $n_T$ are then very different according to the two consistency relations. For this fiducial model, future experiments will then be able to verify the MG consistency relation and rule out the standard consistency relation. We refer readers to Sec. \ref{subsection-forecast-cons-rel} for some details.

It is possible to test the MG consistency relation, Eq. \eqref{eq-MG-consistency-rel}, with future CMB experiments, because $\nu_0$ affects the CMB B-mode power spectrum. We will explore these effects in Sec. \ref{subsection-effect-friction-speed}. If we are able to obtain the values of $\nu_0$, $r$ and $n_T$ from observations, we can then test whether Eq. \eqref{eq-MG-consistency-rel} is satisfied. However, we note that it is possible to do so with CMB data only if $\nu_0$ is constant throughout the history of the Universe, or at least from inflation to recombination. Only in this case, it will be the same MG friction parameter $\nu_0$ in Eq. \eqref{eq-MG-consistency-rel} that also affects the CMB B-mode power spectrum. The value of $\nu_0$ inferred from CMB data is actually the one after inflation (let us call it $\nu_{0,cmb}$), while the $\nu_0$ in the MG consistency relation Eq. \eqref{eq-MG-consistency-rel} is the one during inflation (let us call it $\nu_{0,inf}$). If $\nu_{0,cmb}\neq\nu_{0,inf}$, it will be incorrect to test the MG consistency relation $n_T=-3\nu_{0,inf}-r/8$ with CMB data which only give $\nu_{0,cmb}$. For example, if $\nu_{0,inf}=0$ but $\nu_{0,cmb}\neq0$, the standard consistency relation is correct but we will see a nonzero $\nu_{0,cmb}$ from future CMB experiments. Another example is if $\nu_{0,inf}\neq0$ but $\nu_{0,cmb}=0$, the MG consistency relation is correct but we will not see any extra friction effects from CMB data. Fortunately, even if $\nu_0$ changes its value after inflation, we can still test the standard inflation consistency relation in GR. Indeed, a nonzero $\nu_{0,inf}$ during inflation still breaks the relation between $n_T$ and $r$  in Eq. \eqref{eq-slow-roll-consistency-rel}. If the standard consistency relation is not satisfied by future CMB experiments, one can draw a conclusion  that either GR needs to be modified or the slow-roll inflation theory is inconsistent. In this work, we will assume, for simplicity, that $\nu_0$ is constant.

We will close the section with a brief discussion of possible generalizations of the result of Eq. \eqref{eq-tensor-spectrum}. For example, the result can be generalized to include a constant modified speed parameter $\varepsilon_0$ in addition to a constant friction parameter $\nu_0$. In this case, equation \eqref{eq-tensor-spectrum} can be easily generalized to
\begin{equation}\label{eq-tensor-spectrum-general}
 |h_k^0|^2 = \frac{G(2H)^{2(1+\tilde{\nu_0})}\left[\Gamma(\frac{3}{2}+\tilde{\nu_0})\right]^2}{\pi^3\cdot \big(\sqrt{(1+\varepsilon_0)}\times k\big)^{3+2\tilde{\nu_0}}}~.
\end{equation}
In other words, we have replaced $k$ in Eq. \eqref{eq-tensor-spectrum} with $\sqrt{(1+\varepsilon_0)}\times k$ to obtain Eq. \eqref{eq-tensor-spectrum-general}. But this does not change the dependence of $|h_k^0|^2$ on $k$, which means the tensor spectral index $n_T$ does not depend on a constant modified speed of gravitational waves. So the consistency relation will not be changed due a modified constant speed of gravitational waves.
Additionally, since the wave-propagation equation \eqref{eq-infla1} is a differential equation in time, mathematically the result \eqref{eq-tensor-spectrum-general} can be generalized to cover cases where $\nu_0$ and $\varepsilon_0$ are functions of the comoving wave number $k$. The only difference for such general cases will be that $\nu_0$ and $\varepsilon_0$ in Eq. \eqref{eq-tensor-spectrum-general} become $k$ dependent. But such generalization is not physically meaningful because the function $g$ in the friction term (and hence $\nu_0$) is $k$ independent, and the dispersion relation usually depends on the physical wave number $k/a$ instead of the comoving wave number $k$.

\section{Effects of Tensor mode Modified Gravity Parameters}\label{section-effect}
After investigating the primordial fluctuation during inflation (only for the cases of constant $\nu_0$ and $\varepsilon_0$), the next step is to see how the MG parameters change the evolution of tensor-mode perturbations at later times, and use observational data to put constraints on our MG parameters. In order to do so, we used a modified version of \textsc{camb} \cite{Camb} and \textsc{CosmoMC} \cite{Cosmomc}. In addition to the changes to the scalar sector in \textsc{ISiTGR}, we add modifications of the wave-propagation equation in the tensor sector. For the scalar modes, we refer the modifications of these to packages \textsc{ISiTGR} \cite{ISiTGR-cite1,ISiTGR-cite2}. We add to the top of these modifications the tensor modes.

We already mentioned in Sec. \ref{section-parametrization} some of the constraints on the dispersion relation in the literature. In particular, a subluminal phase speed of gravitational waves is almost forbidden by consideration of gravitational Cherenkov radiation. But, in this work we will not use those as prior bounds but rather aim to obtain independent and complementary constraints. We will constrain our MG parameters solely from the current CMB observations. Our results should thus serve as independent constraints on the dispersion relation. However, some physical ranges need to be imposed on the MG parameters for the stability of the solutions of the perturbation equations:
\begin{enumerate}
  \item $\nu_0>-1$. If not, the friction term in Eq. \eqref{eq-Mo.propagation} has an enhancing instead of suppressing effect.
  \item $\varepsilon_0>-1$. If $\varepsilon_0<-1$, $\omega_T^2=(1+\varepsilon_0)\times\tfrac{k^2}{a^2}$ is negative and tensor modes will all be unstable. We also exclude the situation $\varepsilon_0=-1$ for a practical reason. If $\varepsilon_0=-1$, $h_k=$ constant is a solution of Eq. \eqref{eq-Mo.propagation}. Then tensor modes will not contribute to CMB temperature anisotropy or polarization spectra, and the tensor-to-scalar ratio $r$ can be arbitrarily large. Our allowed range of $\varepsilon_0$ means that we are also considering subluminal phase speeds of gravitational waves (i.e., for $-1<\varepsilon_0<0$).
  \item $(\varepsilon_l)^{n}\geq0$. If not, the squared graviton mass $\mu^2=(\varepsilon_l)^n\times\mu_0^2$ is negative. Tensor modes become tachyonic, and $\omega_T^2$ will be negative for large-scale modes with $k^2/a^2<|\mu^2|$. The evolution of these modes will then grow exponentially and become unstable.
  \item $\varepsilon_h\geq0$. If not, $\omega_T^2$ will be negative for small-scale modes with $k^2/a^2>|\varepsilon_h|\times K_0^2$.
\end{enumerate}

Those physical ranges of MG parameters are also listed in Table \ref{table-all-parameters}.

\subsection{Analyzing the effects of modified friction and nonstandard speed}\label{subsection-effect-friction-speed}
\begin{figure}[tbp]
\includegraphics[width=0.48\textwidth]{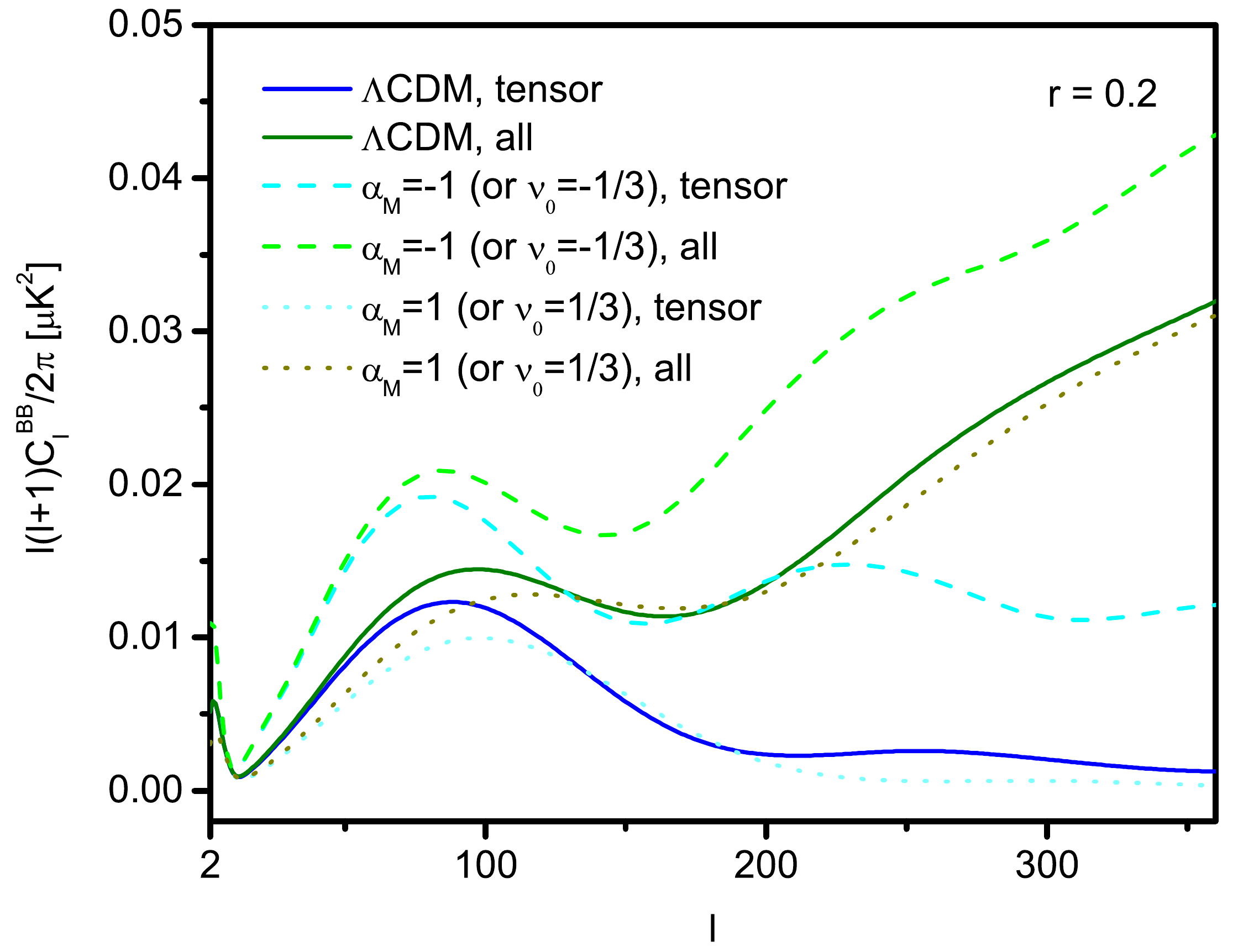}
\caption{\label{fig-REf-camb} Reproducing Fig. 1 from Ref. \cite{friction}. Within the figure,``tenso'' refers to the B-mode due to tensor modes only, and ``all'' includes the lensing in the scalar mode. Notice that we set $r=0.2$ here to reproduce consistent results with Ref. \cite{friction}. Larger friction leads to a smaller tensor-mode amplitude and consequently a smaller tensor-induced B-mode polarization.}
\end{figure}

\begin{figure*}[!htbp]
\includegraphics[width=\textwidth]{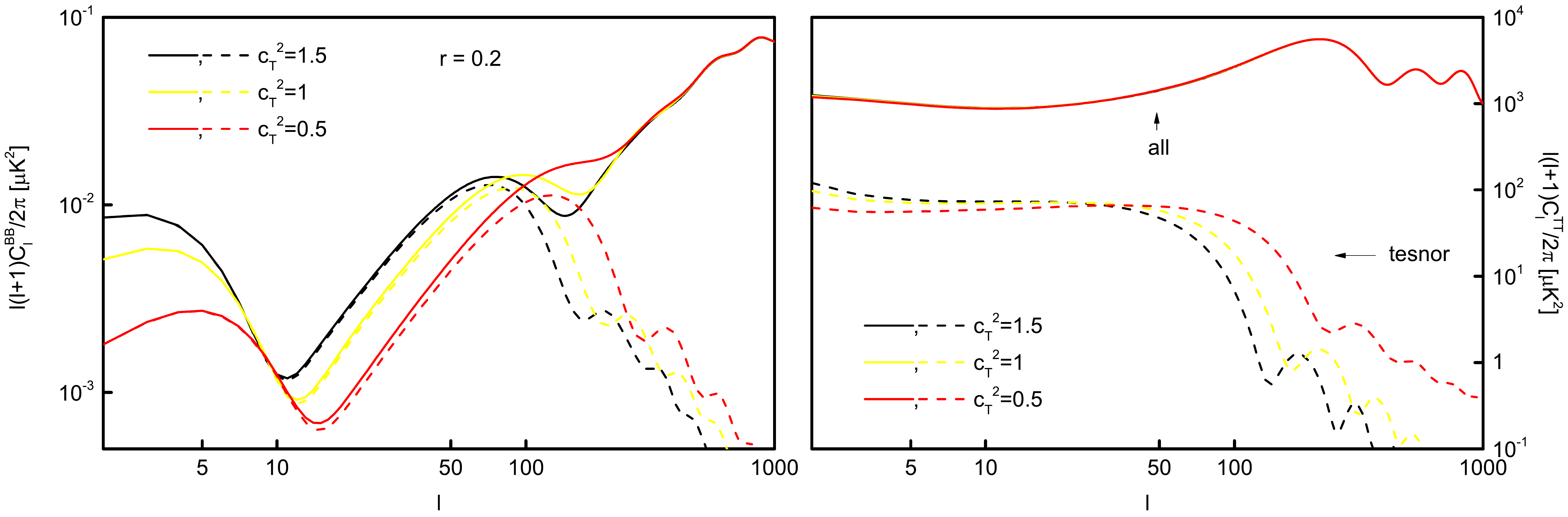}
\caption{\label{RECT2-camb} Reproducing Fig. 1 from Ref. \cite{MG.Ct.of.GW}. We also set $r=0.2$ here to get the same results as Ref. \cite{MG.Ct.of.GW}. In the left panel, we show the effects on the B-mode polarization. The solid lines represent the results due to tensor modes plus lensing, and the dashed lines represent tensor modes only. As explained in Ref. \cite{MG.Ct.of.GW}, modifying the speed of gravitational waves shifts the peaks of the B-mode polarization. The effects on the temperature power spectrum are shown in the right panel. The solid and the dashed lines have the same correspondences as in the left panel. We can see that even if the tensor-induced temperature power spectrum is changed, the total temperature power spectrum is not affected because the scalar modes are dominating.}
\end{figure*}

In this subsection, we explore the effects of the MG parameters $\nu_0$ and $\varepsilon_0$ on the CMB B-mode polarization power spectrum. We vary each one of them individually, and set the other MG parameters to their GR value. To verify our modification in \textsc{camb}, in Fig. \ref{fig-REf-camb} and Fig. \ref{RECT2-camb} we reproduced two figures from Refs. \cite{friction} and  \cite{MG.Ct.of.GW}.

Figure \ref{fig-REf-camb} shows the effects due to different values of $\nu_0$, corresponding to different strengths of friction. In Fig. \ref{fig-REf-camb} we have used $\alpha_M$ to denote the friction term instead of $\nu_0$, in order to be consistent with Ref. \cite{friction}. For the rest of this paper, we use our notation $\nu_0$. Again, for constant $\nu_0$ and $\alpha_M$, they are only different by a factor of $\frac{1}{3}$, and $\nu_0=\frac{1}{3}\alpha_M$. We refer readers to Ref. \cite{friction} for more a detailed analysis of the friction term. For a brief discussion, we can see that a larger $\nu_0$ (or $\alpha_M$) means a larger damping effect, and generally leads to a smaller tensor-mode amplitude. But we need to keep in mind that, a smaller tensor-mode amplitude does not necessarily mean a smaller B-mode polarization induced by tensor-mode perturbations, since it is the time derivative of the amplitude that is important, see Chap. 7 in Ref. \cite{S.W.cosmo.}. However, it turns out in this case that a larger $\nu_0$ (or $\alpha_M$) simply leads to a smaller B-mode, as shown in Fig. \ref{fig-REf-camb}.

Figure \ref{RECT2-camb} shows the effects due to different values of $\varepsilon_0$, corresponding to different speeds of gravitational waves. We do not restrict our parameter $\varepsilon_0$ to be non-negative, which means we do not use the constraint set by the consideration of gravitational Cherenkov radiation, in order to derive complementary results as we explained at the beginning of Sec. \ref{section-effect}. A detailed analysis of a nonstandard speed was given in Ref. \cite{MG.Ct.of.GW}, in which the speed was parametrized as $c_T^2$. Their parametrization is the same as our $1+\varepsilon_0$ parametrization. The major effect of a different $\varepsilon$ is a horizontal shift of the peaks in the B-mode power spectrum. The reason for such peak shifting can be understood as follows. Roughly speaking, for a nonzero $\varepsilon_0$, solutions of Eq. \eqref{eq-Mo.propagation} are changed so that $h_k\rightarrow h_k'=h_{\sqrt{1+\varepsilon_0}k}$. For the same $k$, the frequency (in time) $\omega_T=k/a$ is now replaced by $\omega_T=\sqrt{1+\varepsilon_0}\times k/a$. Consequently, for the same frequency $\omega_T$, the corresponding comoving wave number is now $k/\sqrt{1+\varepsilon_0}$ instead of $k$. If the original peak is at a multiple of $\ell$, it will be shifted to $\frac{\ell}{\sqrt{1+\varepsilon_0}}$. For example, the B-mode recombination peak in GR is around $\ell\sim100$. For $1+\varepsilon_0=1.5$ and $0.5$, this peak will be shifted to $\ell\sim80$ and $\sim140$ respectively, as shown in Fig. \ref{RECT2-camb}. Another effect from a nonstandard speed involves the amplitude of the reionization peak. We can see in Fig. \ref{RECT2-camb} that a smaller speed leads to a smaller amplitude of this peak, in addition to a horizontal shift. This is because a smaller speed makes all modes reenter the horizon later, so that the largest-scale modes remain constant for a longer time and do not contribute to the B-mode production (recall again that the important part is the time derivative of the tensor-mode amplitude). Such a contribution is important for the reionization peak, and so a smaller speed leads to a smaller peak. Vice versa, a larger speed makes the largest-scale modes reenter the horizon, and oscillate earlier and participate in the B-mode production. 

\subsection{Effects of large-scale deviation}\label{subsection-mass-effect}
The large-scale (low-$k/a$) deviation represents a constant graviton mass. Again, the squared mass $\mu^2$ needs to be non-negative to avoid small-scale tachyonic instability. If $\mu^2$ is negative, roughly speaking the solution will grow exponentially for the modes with $k^2/a^2+\mu^2<0$.

\begin{figure*}[!t]
\includegraphics[width=\textwidth]{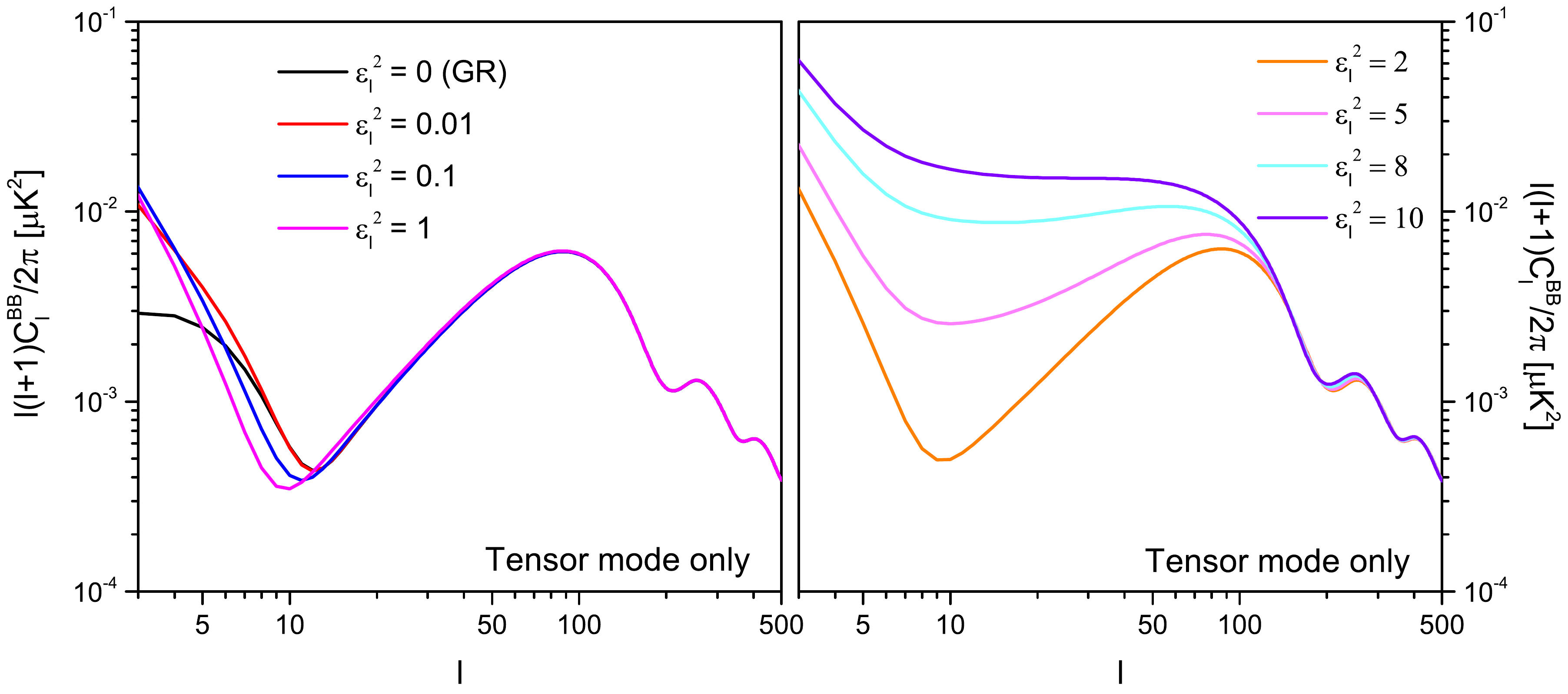}
\caption{\label{Re-massive} The effects of the large-scale deviation on the tensor-induced B-mode polarization. Both panels have the same horizontal and vertical scales. In the left panel, for a small $\varepsilon_l$, a larger $\varepsilon_l$ leads to a smaller large-scale B-mode polarization. In the right panel,the opposite effects take place. For a large $\varepsilon_l$, a larger $\varepsilon_l$ leads to greater a large-scale B-mode polarization. These results are consistent with those in Ref. \cite{MassiveGinCMB}, where we can see that the amplitude of the tensor-induced B-mode has an oscillating dependence on the graviton mass $\mu$. See the text for a discussion.}
\end{figure*}

An analysis of the effects on the CMB due to a graviton mass has been given in Ref. \cite{MassiveGinCMB}. The authors there estimated an upper bound of the graviton mass, $\mu\lesssim10^{-30}\,eV$, for a nonvanishing tensor-to-scalar ratio. Here we reproduce some of their numerical results and show them in Fig. \ref{Re-massive}. A similar upper bound of the graviton mass will be obtained in Sec. \ref{subsection-constraint-mass}, where, instead of estimating, we will use a MCMC analysis and get constraints from the current available data. In Fig. \ref{Re-massive}, since the effects are not monotonic with  $\varepsilon_l$, we show them in two panels. In fact, the effects have an oscillating dependence on $\varepsilon_l$, as we will explain in the next paragraph. We only show the effects on the B-mode polarization, because the temperature and E-mode are dominated by the scalar modes.

Depending on the time ordering of recombination, the horizon reentering (when $k/a\sim H$), and the transition from being relativistic to nonrelativistic (when $k/a\sim \mu$), there are different effects on the evolutions of different perturbation modes. We can qualitatively see that as follows. With a finite graviton mass, there is a distinct feature from GR for the perturbation evolutions: all perturbation modes will eventually become nonrelativistic (i.e., $k/a<\mu$, or the momentum of a graviton is smaller than its mass). Since the physical wave numbers decrease with time, perturbation modes always start out being relativistic (i.e, $k/a>\mu$), and later transition to nonrelativistic (i.e, $k/a<\mu$). And once they become nonrelativistic, they remain so. The time for the relativistic-to-nonrelativistic transition is roughly determined by the condition $k/a\sim \mu$, which depends on $k$. Different modes have different transition times. Consider only the polarizations produced near recombination: for the modes whose relativistic-to-nonrelativistic transitions happen after recombination (true for small-scale modes), their evolutions before recombination will be almost the same as in GR. Therefore, their contributions to the CMB temperature and polarization will be nearly unchanged. For the modes whose transitions happen before recombination, the situation is different and interesting effects take place, but the analysis will be more involved. Detailed discussions were provided in Ref. \cite{MassiveGinCMB}, in which perturbation modes were divided into three classes: class I consists of modes that are relativistic at recombination; class II consists of modes that are nonrelativistic as they enter the horizon; and class III consists of modes that are relativistic when they reenter the horizon and become nonrelativistic during recombination. Depending on whether the graviton mass is larger or smaller than the Hubble rate at recombination, the third class may or may not exist.

Now we discuss whether the largest-scale  modes (small wave number compared to $\mu$ and $H$) are well behaved for a finite $\mu^2$. The discussion here will also explain the oscillatory dependence of the large-scale effects. Consider the largest-scale modes with $k/a$ negligible compared to $\mu$ and $H$. In this simple situation, Eq. \eqref{eq-Mo.propagation} becomes,
\begin{equation}\label{large-mass}
\ddot{h}_{k}+\frac{2}{t}\dot{h}_k+\mu^2h_k=0.
\end{equation}
Solutions to Eq. \eqref{large-mass} are the spherical Bessel functions of order $0$. The asymptotically constant initial condition gives,
\begin{equation}\label{eq-large-modes-massive}
h_k(t)\propto j_0(\mu t)=\frac{\sin(\mu t)}{\mu t}~,
\end{equation}
where $j_0(x)$ is the spherical Bessel function of the first kind of order $0$. It means that with a finite $\mu$, the largest-scale-mode evolutions do not depend on $k$, and they start to oscillate earlier than they would in GR. So the largest-scale modes are well behaved. If the graviton mass is large enough (more explicitly, larger than the Hubble rate at recombination, i.e., $\mu>H_{recom}$), they oscillate before recombination, and consequently contribute to the CMB temperature anisotropy and polarization spectra. In contrast, in GR, the largest-scale modes remain constant and do not contribute. Since the tensor-mode amplitude has an oscillating dependence on the graviton mass (and hence on $\varepsilon_l$) as shown in Eq. \eqref{eq-large-modes-massive}, the largest-scale-mode contribution to the B-mode polarization in MG also has an oscillating dependence on $\varepsilon_l$. As shown in the left panel of Fig. \ref{Re-massive}, for small $\varepsilon_l$, the low-$\ell$ spectrum of the B-mode polarization decreases with $\varepsilon_l$. But in the right panel, for larger $\varepsilon_l$, it increases with $\varepsilon_l$. A more detailed analysis and similar numerical results were given in Ref. \cite{MassiveGinCMB}, where they showed two more panels, and the B-mode spectrum decreases and increases again with even larger graviton masses.


\subsection{Effects of small-scale deviation}
\begin{figure}[!t]
\includegraphics[width=0.48\textwidth]{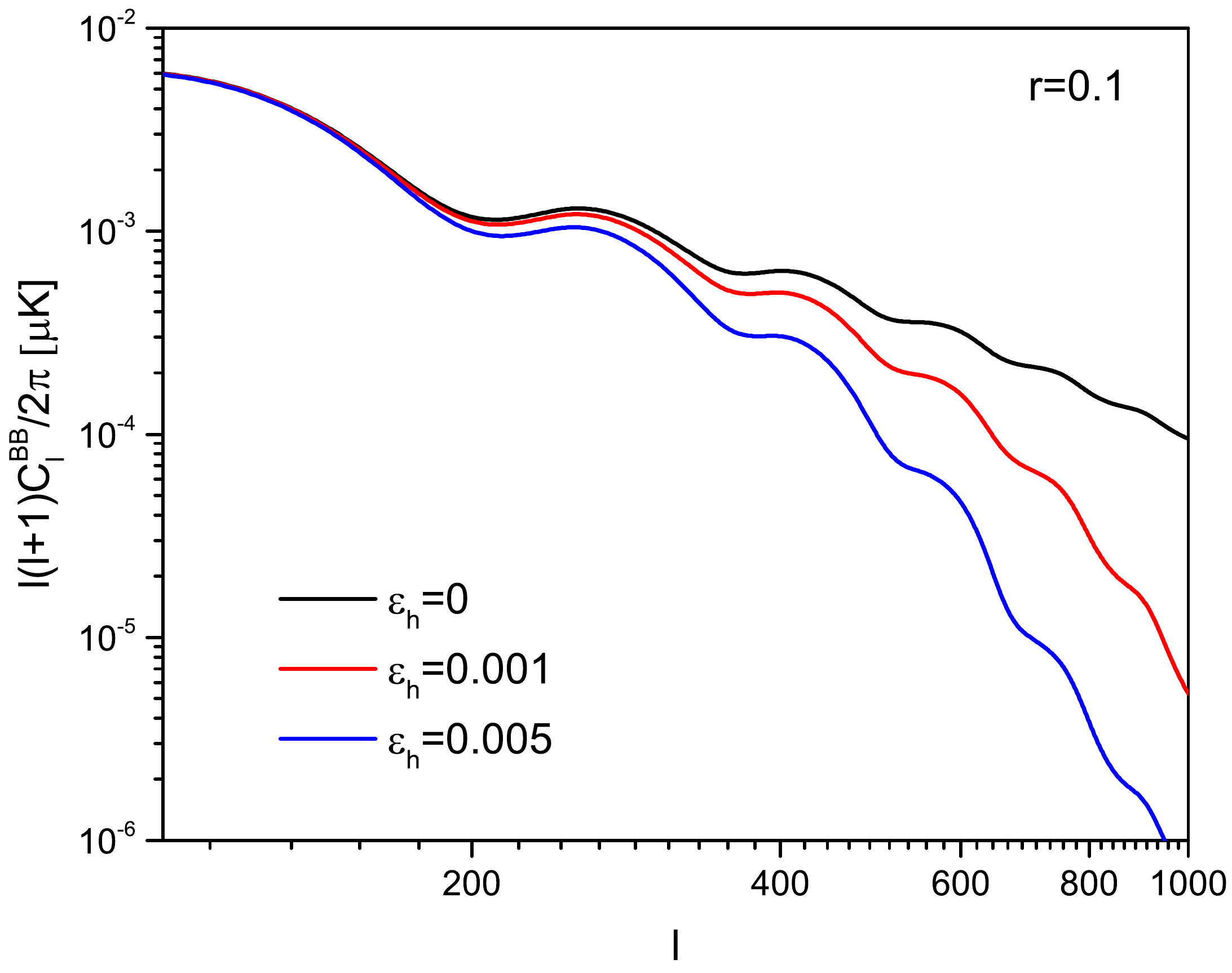}
\caption{\label{fig-high-camb} Effects of small-scale (high-$k/a$) deviation on the B-mode power spectrum. Here we only show the tensor-induced B-mode polarization. The spectrum at small scales (low $\ell$) is not affected as expected. A larger $\varepsilon_h$ makes the small-scale modes reenter the horizon earlier, resulting in a smaller tensor-mode amplitude and consequently a smaller B-mode polarization. This effect is hard to observe since the dominating B-mode polarization at small scales is from the lensed E-mode.}
\end{figure}
In this subsection we investigate the effects of the small-scale (high-$k/a$) parameter $\varepsilon_h$ on the B-mode polarization. Figure \ref{fig-high-camb} shows the results of the B-mode polarization power spectrum for different values of $\varepsilon_h$. Here we set $r=0.1$. Recall that we restrict $\varepsilon_h$ to be non-negative because a negative $\varepsilon_h$ can lead to small-scale instability. This small-scale instability can be seen from Eq. \eqref{eq-dispersion-a} and Eq. \eqref{eq-dispersion-cases}, and when $\varepsilon_h\left(\frac{k/a}{K_0}\right)^2<-1$ the squared frequency $\omega_T^2$ becomes negative. If one wants to allow a negative $\varepsilon_h$, it is necessary to introduce a cutoff or include a positive higher-order term. We will not do these, because, first, the cutoff is totally arbitrary and the results are not converging for higher and higher cutoffs. A higher cutoff only leads to a higher amplitude. Second, to include a positive higher-order term requires another parameter specifying the physical wave number from which the higher-order term becomes significant. Doing so requires more complicated considerations, such as analyzing the competition of the second-order term and the higher-order term. So for simplicity we keep the number of parameters to be a minimum, but we are still be able to catch some (if not most) of the physics of modified gravity at small scales.

\begin{figure*}[!tp]
\includegraphics[width=0.495\textwidth]{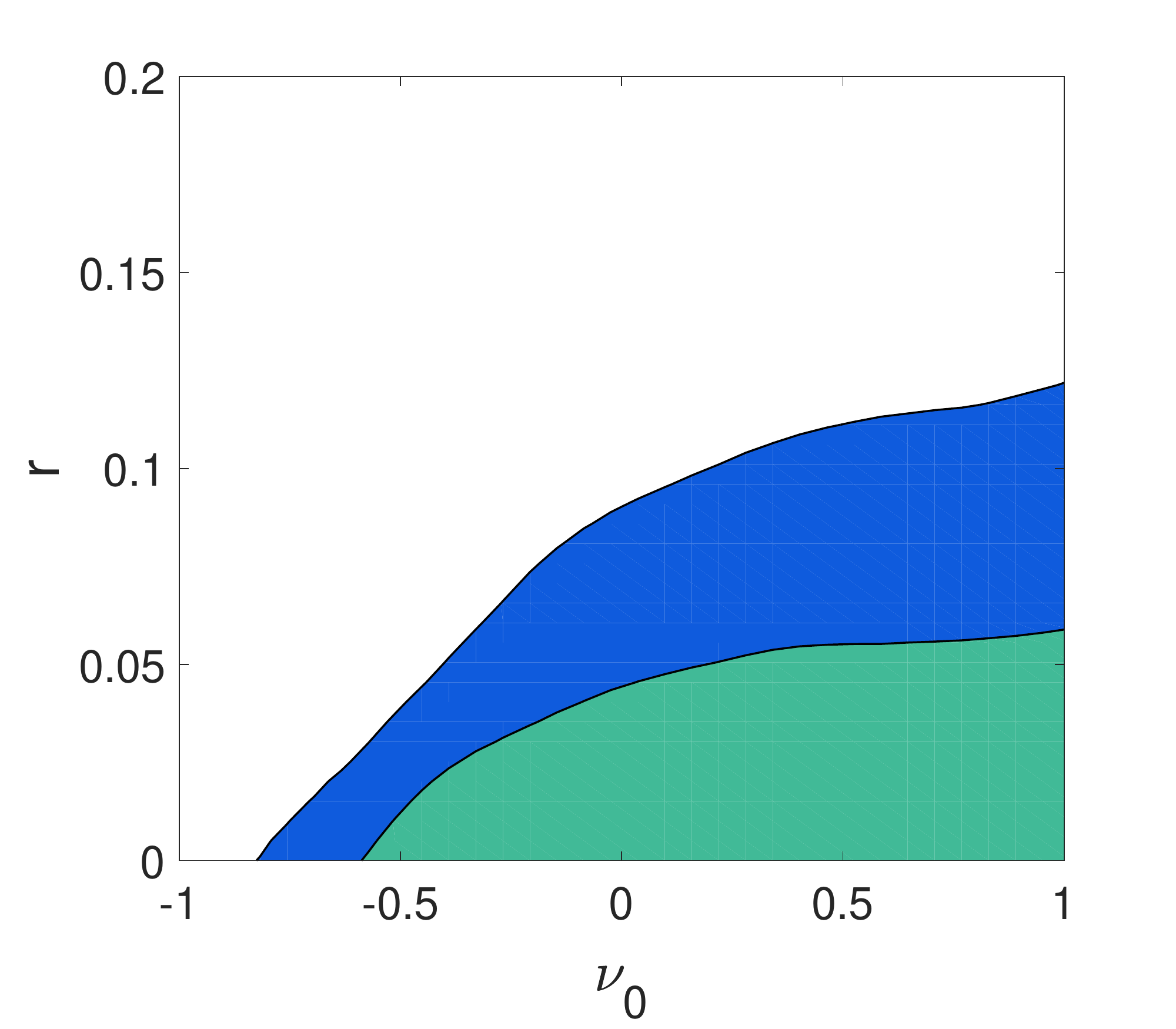} 
\includegraphics[width=0.495\textwidth]{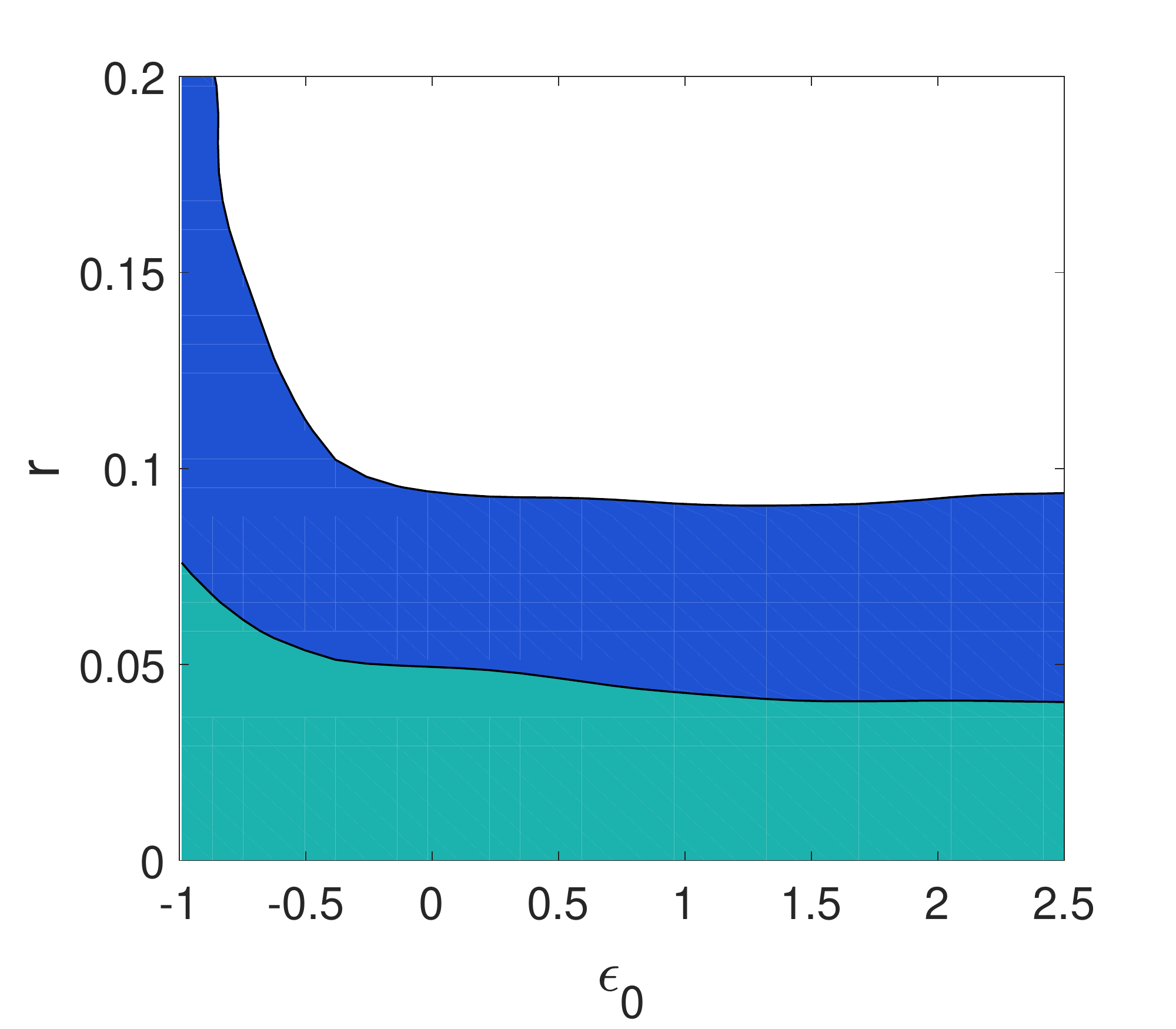}
\caption{\label{nu0-epsilon0-plc15-BKP}The $1$-$\sigma$ (green) and $2$-$\sigma$ (blue+green) confidence levels of marginalized constraints in the $r$ vs $\nu_0$ (left panel) and the $r$ vs $\varepsilon_0$ (right panel) parameter spaces. Equivalently, we can say the white parameter region is disfavored at the $95\%$ confidence level.}
\end{figure*}

As Fig. \ref{fig-high-camb} shows, the tensor-induced B-mode polarization power spectrum can be significantly suppressed at small scales (large $\ell$) while keeping it unaffected at large scales (small $\ell$), as expected. The effects of small-scale deviation can be understood as follows. A nonzero $\varepsilon_h$ changes the time of horizon reentering. For a certain mode with comoving wave number $k$, a larger $\varepsilon_h$ leads to earlier horizon reentering, resulting in a smaller tensor-mode amplitude. So the tensor-induced B-mode is expected to be smaller.

This small-scale deviation is difficult to observe, because it hardly changes the total B-mode power spectrum at small scales, where the contribution from lensing is dominating. A larger $\varepsilon_h$ only makes the tensor-mode contribution less significant in the high-$\ell$ spectrum. Consequently, the dominating B-mode from lensing at small scales makes it very difficult to set a constraint on the parameter $\varepsilon_h$. So we will not do the corresponding Monte Carlo analysis for $\varepsilon_h$ and leave it for future data. Fortunately, with the near-future CMB experiments we will be able to see such small-scale effects, if $\varepsilon_h$ is large enough so that small-scale deviation begins with a large-enough-scale onset. We will estimate the constraint on $\varepsilon_h$ with the Fisher matrix formalism in Sec. \ref{section-forecast}.

\section{Constraints on Tensor Mode Modified Gravity Parameters}\label{section-constraints}
Tensor-mode perturbations, if present, can smooth out the temperature-anisotropy power spectrum and generate E-mode and B-mode polarization patterns in the CMB. Therefore, both CMB temperature and polarization maps can be used to constrain the parameters related to tensor-mode perturbations. In the following subsections, we study the constraints on the four MG parameters individually. For example, when we are constraining $\nu_0$, we fix $\varepsilon_0$, $\varepsilon_h$ and $\varepsilon_l$ to their GR values. We do that for a practical reason since current data gives very weak constraints on the tensor-mode MG parameters. It is computationally expensive to constrain the MG parameters simultaneously. In the MCMC analysis, we also fix the six standard cosmological parameters to the values of the Planck 2015 best fit \cite{Planck2015XIII-Cos.Param.}, and constrain the tensor-to-scalar ratio $r$ with one of the tensor-mode MG parameters at a time using the joint data of Planck and BICEP2 \cite{BKP2014} and the Planck 2015 low-$\ell$ polarization data \cite{Planck2015XIII-Cos.Param.}. In this section, we use the standard inflation consistency relation on the value of $n_T$, namely, $n_T=-r/8$. For the current data, we will not vary the tensor spectral index $n_T$ since otherwise the parameter space would be too large and give no useful information.

For current data, the tensor-induced B-mode polarization has not been detected yet so we will provide only some bounds on the MG parameters. Due to the weak constraining power of current data, we will also not attempt any joint constraints on the four MG parameters. We also do not constrain $\varepsilon_h$ because the observed high-$\ell$ B-mode polarization is dominated by the lensed E-mode, so current data only give a large and meaningless allowed region in the $r$ vs $\varepsilon_h$ parameter space. Instead, we will forecast the constraint on $\varepsilon_h$ in Sec. \ref{section-forecast} for some future experiments.

\subsection{Updating the constraints on friction and constant speed using the new BKP data}\label{subsection-constraint-friction-speed}
We first update the constraints on the friction and the speed by using the data from the Planck-BICEP2 joint analysis (BKP)  \cite{BKP2014} and the Planck 2015 low-$\ell$ polarization data \cite{Planck2015XIII-Cos.Param.}. To validate our modification to \textsc{camb}, we reproduced the marginalized likelihood distributions in the $\alpha_M$ vs $r$ and $r$ vs $c_T^2$ parameter spaces in Ref. \cite{friction} using the old BICEP2 data \cite{BICEP2I}, and we got the same results.

The left panel in Fig. \ref{nu0-epsilon0-plc15-BKP} shows the marginalized constraints in the $r$ vs $\nu_0$ parameter space using the BKP and the Planck 2015 low-$\ell$ polarization data. The black curves are iso-likelihood contours, within which the integrated probabilities are $68\%$ and $95\%$ respectively. Consequently, the green and the blue+green regions respectively correspond to the $1$-$\sigma$ ($68\%$) and $2$-$\sigma$ ($95\%$) confidence levels (C.L.). There is a probability of $68\%$ for the true values of $r$ and $\nu_0$ to be located within the green region, and $95\%$ within the blue+green region. In other words, at the $95\%$ C.L., the white parameter space is ruled out. (Note that the blue-only region is ruled out at the $68\%$ C.L., but allowed at the $95\%$ C.L.). We can see from the left panel of Fig. \ref{nu0-epsilon0-plc15-BKP} that the degenerate direction goes roughly as $r-0.05\nu_0=$ constant, consistent with that in Ref. \cite{friction}. The tensor-to-scalar ratio $r$ is consistently zero. We cut out the large $\nu_0$ parameter space, because a larger $\nu_0$ only leads to a larger allowed tensor-to-scalar ratio $r$.

Using the same data, in the right panel of Fig. \ref{nu0-epsilon0-plc15-BKP} we show the constraints in the $r$ vs $\varepsilon_0$ parameter space. The green and blue regions have the same meanings as those in the left panel of Fig. \ref{nu0-epsilon0-plc15-BKP}. Since we have not observed the tensor-induced B-mode polarization, we should not expect the peak position of the B-mode power spectrum to constrain the speed of gravitational waves as in Ref. \cite{MG.Ct.of.GW}. Instead we see in the right panel of Fig. \ref{nu0-epsilon0-plc15-BKP} that a smaller $\varepsilon_0$ (and hence a smaller speed) allows a larger tensor-to-scalar ratio. As $\varepsilon_0$ approaches $-1$, at the $1$-$\sigma$ C.L., we have an upper limit of $r$ $\sim1.75$ shown by the green region in the right panel of Fig. \ref{nu0-epsilon0-plc15-BKP}. As mentioned in Sec. \ref{section-effect}, a smaller speed means a later horizon reentering. An extreme case is a vanishing speed ($\varepsilon_0 = -1$), in which the tensor-mode perturbations would never reenter the horizon and their amplitudes would always remain constant. Since the tensor-induced B-mode polarization requires time variation of the tensor-mode perturbations, a vanishing speed then means no tensor-induced B-mode polarization and $r$ can be arbitrarily large. This is also why we excluded the parameter value $\varepsilon_0 = -1$ in the MCMC analysis. The arbitrarily large value of the allowed $r$ as $\varepsilon_0$ approaches $-1$ is shown by the blue region in the right panel of Fig. \ref{nu0-epsilon0-plc15-BKP}. On the other hand, larger $\varepsilon_0$ does not seem to affect the constraint on $r$ very much. This is because, besides making the tensor-mode amplitudes vary with time, horizon reentering also makes them smaller. A larger $\varepsilon_0$ then has both an enhancing effect (due to the time-varying tensor-mode amplitudes) and a suppressing effect (due to smaller amplitudes) on the CMB B-mode polarization.

\begin{figure}[tbp]
\includegraphics[width=0.48\textwidth]{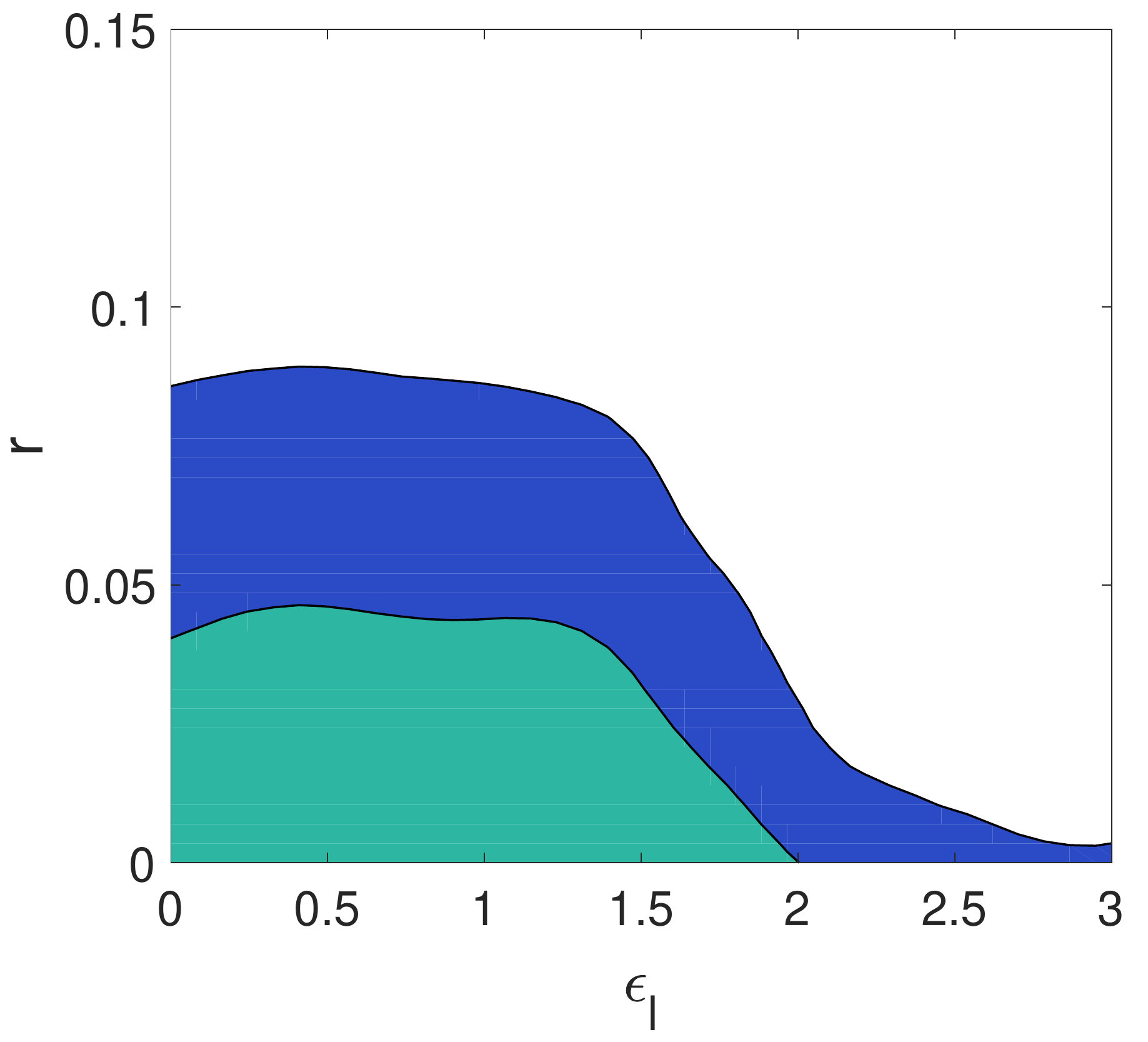}
\caption{\label{fig-mass-like} Constraints in the $r$ vs $\varepsilon_l$ parameter space. The plateau from $\varepsilon_l=0$ to $\sim1.5$ means this range of $\varepsilon_l$ makes little difference on the constraint of $r$, which is similar to the massless case. Unless $r$ is very small, the sharp drop of the  allowed value of $r$ after $\varepsilon_l\sim1.5$ sets an upper bound of the graviton mass, $\mu_{upper}\sim1.4\times10^{-29}\,eV$, for most allowed values of $r$.}
\end{figure}
\subsection{Constraints on large-scale deviation}\label{subsection-constraint-mass}
Using the same data, we obtained the constraints in the $r$ vs $\varepsilon_l$ parameter space as shown in Fig. \ref{fig-mass-like}. The conversion between $\varepsilon_l$ and the graviton mass $\mu$ [for $n=4$ in Eq. \eqref{eq-dispersion-cases}] is $\mu=\varepsilon_l^2\times5.238\times10^{-58}\,M_p=\varepsilon_l^2\times6.395\times10^{-30}\,eV$. We can see that the constraint of $r$ is insensitive to the parameter $\varepsilon_l$ for $\varepsilon_l\lesssim1.5$, which means a graviton mass smaller than $\sim10^{-29}\,eV$ should have no observational effect on the CMB for the current level of sensitivity. The constraint of $r$ in this range of $\varepsilon_l$ is roughly the same as the case in GR. Both the $1$-$\sigma$ and $2$-$\sigma$ contours have relatively sharp turns at $\varepsilon_l\sim1.5$. A larger $\varepsilon_l$ leads to significant drops of the allowed value of $r$ for both contours. This location ($\varepsilon_l\sim1.5$) of the sharp turns roughly corresponds to an upper bound of the graviton mass $\mu_{upper}\sim1.4\times10^{-29}\,eV$ unless $r$ is very small. This upper bound is roughly of the same order of magnitude a the estimation in Ref. \cite{MassiveGinCMB}. Note that, if massive gravity is responsible for the late-time cosmic acceleration, the graviton mass should be of the order of the Hubble constant $H_0$ (in natural units) \cite{MassiveGinCMB,MassiveGravity2014}, which is $\sim10^{-33}\,eV$ and is about $3\sim4$ orders of magnitude smaller than the rough upper bound (for nonvanishing $r$) obtained in this work.

\begin{table*}[!tpb]
\caption{\label{COrEspec}Specifications of the COrE mission obtained from Ref. \cite{COrEWhitePaper}. $f_{sky}=0.7$. Here, $\nu$ denotes the central frequency of each band, (not our friction parameter).}
\begin{ruledtabular}
\begin{tabular}{lccccccccccccccc}
$\nu/({\rm{{\rm{GHz}}}})$ & $45$ & $75$ & $105$ & $135$ & $165$ & $195$ & $225$ & $255$ & $285$ & $315$ & $375$ & $435$ & $555$ & $675$ & $795$ \\
\hline
$\Delta\nu/({\rm{{\rm{GHz}}}})$ & $15$ & $15$& $15$& $15$& $15$& $15$& $15$& $15$& $15$& $15$& $15$& $15$& $195$& $195$& $195$ \\
$\theta_{fwhm}/(\rm{arcmin})$ & $23.3$ & $14.0$ & $10.0$ & $7.8$ & $6.4$ & $5.4$ & $4.7$ & $4.1$ & $3.7$ & $3.3$ & $2.8$ & $2.4$ & $1.9$ & $1.6$ & $1.3$\\
 Pol. RJ &\multirow{2}{*}{ $8.61$} & \multirow{2}{*}{$4.09$} & \multirow{2}{*}{$3.50$} & \multirow{2}{*}{$2.90$} & \multirow{2}{*}{$2.38$} & \multirow{2}{*}{$1.84$} & \multirow{2}{*}{$1.42$} & \multirow{2}{*}{$2.43$} & \multirow{2}{*}{$2.94$} & \multirow{2}{*}{$5.62$} & \multirow{2}{*}{$7.01$} & \multirow{2}{*}{$7.12$} & \multirow{2}{*}{$3.39$} & \multirow{2}{*}{$3.52$} & \multirow{2}{*}{$3.60$}\\
$(\mu \rm{K}\cdot \rm{arcmin})$& & & & & & & & & & & & & & &\\
\end{tabular}
\end{ruledtabular}
\end{table*}

\begin{table}
\begin{ruledtabular}
\caption{\label{S4spec}Specifications of Stage-IV obtained and calculated from Ref. \cite{CMB-Bmode-forecast2015}. $f_{sky}=0.5$.}
\begin{tabular}{lccccc}
$\nu/({\rm{{\rm{GHz}}}})$ & $40$ & $90$ & $150$ & $220$ & $280$ \\
\hline
$\Delta\nu/({\rm{{\rm{GHz}}}})$ & \multicolumn{5}{c}{$30\%$ fractional bandpass}\\
$\theta_{fwhm}/(\rm{arcmin})$ & $11.0$ & $5.0$ & $3.0$ & $2.0$ & $1.5$\\
 Pol. RJ ($\mu \rm{K}\cdot \rm{arcmin}$) & $2.9$ & $1.2$ & $0.86$ & $1.6$ & $1.6$\\
\end{tabular}
\caption{\label{PIXIEspec}Specifications of PIXIE obtained from Ref. \cite{PIXIE2011}. $f_{sky}=0.7$.}
\begin{tabular}{lccm{0.22\textwidth}}
$\nu/$ {\rm{{\rm{GHz}}}} & $\frac{\Delta\nu}{({\rm{{\rm{GHz}}}})}$& $\frac{\theta_{fwhm}}{(\rm{arcmin})}$ & \multicolumn{1}{c}{Pol. RJ ($\mu \rm{K}\cdot arcmin$)} \\
\hline
$15:7665$ & $15$ & $96$  & The sensitivities of the $511$ frequency channels are provided by Ref. \cite{private1}.\\
\end{tabular}
\end{ruledtabular}
\end{table}

There is an allowed parameter-space ``tail'' for $\varepsilon_l\gtrsim2.5$. This ``tail'' extends to very large $\varepsilon_l$ which has been cut off in Fig. \ref{fig-mass-like}. This ``tail'' is present because, as $r$ approaches $0$, the amplitude of tensor-mode perturbations approaches $0$ as well. Then there would be no tensor-induced effects on the CMB (temperature or polarization), and $\varepsilon_l$ (and the graviton mass) can be arbitrarily large.

\section{Forecast of constraints on Tensor mode modified gravity parameters}\label{section-forecast}
In this section, we use the Fisher matrix formalism to forecast the constraints on the tensor-mode MG parameters that could be obtained by the COrE mission \cite{COrEWhitePaper}, CMB Stage-IV \cite{Stage-IV-paper1} and PIXIE \cite{PIXIE2011}. Tables \ref{COrEspec}, \ref{S4spec} and \ref{PIXIEspec} list the specifications of these three near-future experiments. To do the forecast correctly, we need to take into account the diffuse foreground components.  Following the method described in Refs. \cite{Spectra-estimation-Bonaldi2006}, we calculate the degraded-noise power spectrum $N_\ell^{post}$ after a component separation. To calculate the foreground residuals, we use the framework described in Ref. \cite{B-mode-forecast-Errard2011,CMB-Bmode-forecast2015}. We include in the analysis the synchrotron and dust as the dominant diffuse foregrounds. So the number of signal components $n_{comp}$ is three including CMB. We denote CMB  as the $0$ component, the synchrotron as $1$ and the dust as $2$.

\subsection{Formalism of CMB forecast and foreground residuals estimation}
With the likelihood provided in Ref. \cite{COrEWhitePaper}, the Fisher matrix reads,
\begin{equation}
\begin{split}
F_{ij}&=-\left<\frac{\partial^2(\ln\mathcal{L})}{\partial\theta_i\partial\theta_j}\right>\\
&= \frac{f_{sky}}{2}\sum_\ell(2\ell+1)Tr\left[\boldsymbol{R_\ell}^{-1}\frac{\partial \boldsymbol{C_\ell}}{\partial \theta_j}\boldsymbol{R_\ell^{-1}}\frac{\partial \boldsymbol{C_\ell}}{\partial\theta_j}\right]~,
\end{split}
\end{equation}
where $\boldsymbol{\theta}$ is the parameter vector of a model, $\boldsymbol{R_\ell}$ is the summation of the theoretical power spectra and the total noise-like power spectra $\boldsymbol{R_\ell}=\boldsymbol{C_\ell}+\boldsymbol{N_\ell^{cmb}}$, where,
\begin{equation}
\begin{split}
&\boldsymbol{C_\ell}=
\begin{pmatrix}
C_\ell^{TT} & C_\ell^{TE} & 0\\
C_\ell^{TE} & C_\ell^{EE} & 0\\
0 & 0 &C_\ell^{BB}
\end{pmatrix}
~,~\\
& {\rm{and}}~\boldsymbol{N_\ell^{cmb}}=
\begin{pmatrix}
N_\ell^{TT} & 0 & 0\\
0 & N_\ell^{EE} & 0\\
0 & 0 &N_\ell^{BB}
\end{pmatrix}~.
\end{split}
\end{equation}
For the B-mode polarization, the theoretical power spectrum is the summation of the contributions from tensor modes and lensing. We do not consider delensing.

Since we are considering foreground subtraction, we take the summation of the degraded (or post-component-separation) noise $N_\ell^{post}$ and the foreground residuals $C_\ell^{fg,res}$ as the total noise-like power spectrum \cite{COrEWhitePaper,CMB-Bmode-forecast2015}. For the B-mode,
\begin{equation}
N_\ell^{BB}=N_\ell^{post}+C_\ell^{fg,res}~.
\end{equation}
The degraded-noise power spectrum is obtained by,
\begin{equation}\label{eq-degraded-noise}
N_\ell^{post}=\Big((\boldsymbol{A}^T\boldsymbol{N_\ell}^{-1}\boldsymbol{A})^{-1}\Big)_{cmb,cmb}~,
\end{equation}
where $\boldsymbol{N_\ell}$ is the instrumental-noise power spectra before component separation, which is assumed to be a $n_{chan}\times n_{chan}$ diagonal matrix for each multiple $\ell$. The diagonal element of $\boldsymbol{N_\ell}$ is given by,
\begin{equation}
\big(\boldsymbol{N_{\ell}}\big)_{\nu\nu}=(\Delta\Omega\sigma_v^2)\exp\left(-\ell(\ell+1)\frac{\theta_{fwhm}^2(\nu)}{8\ln2}\right)~,
\end{equation}
where the index $\nu$ (not our friction parameter) denotes the central frequency of a channel, and there are $n_{chan}$ channels. For example, for the COrE mission, there are $n_{chan}=15$ frequency channels as shown in the first row in Table \ref{COrEspec}. The full-width-at-half-maximum angle $\theta_{fwhm}(\nu)$ and the quantity $\Delta\Omega\sigma_v^2$ (inverse of the weight) can be obtained from the third and the forth rows in Table \ref{COrEspec}. The $n_{chan}\times n_{comp}$ mixing metric $\boldsymbol{A}$ in Eq. \eqref{eq-degraded-noise} is calculated as,
\begin{equation}\label{eq-mixing-matrix}
A_{\nu i}=\int d\nu' \delta_\nu(\nu')A_i^{raw}(\nu')~,
\end{equation}
where the index $i$ can be $cmb$, $sync$ or $dust$, denoting the signal components. Different components can be separated  because they have different emission laws. Different emission laws are expressed as different antenna-temperature functions $A_i^{raw}(\nu')$ of frequency $\nu'$. In Eq. \eqref{eq-mixing-matrix} $\delta_\nu(\nu')$ is a normalized band-pass-filter function for each channel. Take the COrE specification for example: the central frequency $\nu$ and the frequency width $\Delta\nu$ of $\delta_\nu(\nu')$ are given by the first and second rows in Table \ref{COrEspec}. For CMB, the antenna temperature reads,
\begin{equation}\label{eq-antenna-temp-cmb}
A_{cmb}^{raw}(\nu)=\frac{(\nu/T_{cmb})^2\exp(\nu/T_{cmb})}{[\exp(\nu/T_{cmb})-1]^2}~.
\end{equation}
We have set $h=k_B=1$. The temperature of the CMB $T_{cmb}$ is $2.73$ K, corresponding to $56.7 $ {\rm{{\rm{GHz}}}}.

For the synchrotron, the antenna temperature follows a power law,
\begin{equation}\label{eq-antenna-temp-sync}
A_{sync}^{raw}(\nu)\propto\left(\frac{\nu}{\nu_{ref,s}}\right)^{\beta_s}~,
\end{equation}
where the reference frequency $\nu_{ref,s}$ will be set to $30 $ {\rm{{\rm{GHz}}}} to be consistent with that for the Planck 2015 synchrotron polarization map \cite{Planck2015-diffuse-components}. If it is only the CMB component that concerns us, the proportional coefficient in Eq. \eqref{eq-antenna-temp-sync} is irrelevant. Since any other proportional coefficient can be absorbed into a redefined $\nu_{ref,s}$, the value of $\nu_{ref,s}$ is actually also irrelevant when we only care about the CMB component. The estimated synchrotron spectral index $\beta_s$ is $-3.1$.

For the dust, the antenna-temperature function follows a grey-body radiation distribution,
\begin{equation}\label{eq-antenna-temp-dust}
A_{dust}^{raw}(\nu)\propto\left(\frac{\nu}{\nu_{ref,d}}\right)^{\beta_d+1}\left(\frac{\exp\left(\frac{\nu_{ref,d}} {T_d}\right)-1}{\exp\left(\frac{\nu}{T_d}\right)-1}\right)~,
\end{equation}
The dust reference frequency $\nu_{ref,d}=353$ {\rm{GHz}} is chosen to be consistent with the one for the Planck 2015 dust polarization map, but again its value is irrelevant when we only care about the CMB component. The dust temperature $T_d$ is fixed to $19.6$ K \cite{B-mode-forecast-Errard2011}. The estimated dust spectral index is $\beta_d=1.59$. We assume the emission laws for synchrotron and dust are spatially independent.

We follow the framework described in Refs. \cite{B-mode-forecast-Errard2011,CMB-Bmode-forecast2015} to calculate the foreground residuals. The idea is as follows. Since we do not exactly know what emission laws are followed by the synchrotron and the dust, the subtraction of those two components from the signal is not ideal. Assuming that the synchrotron and the dust emission laws take the form of Eq. \eqref{eq-antenna-temp-sync} and Eq. \eqref{eq-antenna-temp-dust}, our uncertainties are now on the two spectral indices $\beta_s$ and $\beta_d$ ($T_d$ is fixed here). One first estimates the uncertainties on the spectral indices $\beta_s$ and $\beta_d$, and then infers the propagated errors in the foreground subtraction. These errors are identified as the foreground residuals. According to Ref. \cite{B-mode-forecast-Errard2011}, the uncertainties of the spectral indices are specified by the matrix $\boldsymbol{\Sigma}$, which is calculated as,
\begin{equation}\label{eq-Sigma-for-beta}
\begin{split}
\Big(\boldsymbol{\Sigma}^{-1}\Big)_{\beta\beta'}=-Tr\Big\{\big[&\frac{\partial\boldsymbol{A}^T}{\partial \beta} \boldsymbol{N}^{-1}\boldsymbol{A} \boldsymbol{C}_N \boldsymbol{A}^T\boldsymbol{N}^{-1}\frac{\partial \boldsymbol{A}}{\partial \beta'}\\
&-\frac{\partial \boldsymbol{A}^T}{\partial \beta} \boldsymbol{N}^{-1}\frac{\partial \boldsymbol{A}}{\partial \beta'}\big]\times\boldsymbol{\hat{F}}\Big\}~.
\end{split}
\end{equation}
where $\boldsymbol{C}_N=(\boldsymbol{A}^T \boldsymbol{N}^{-1} \boldsymbol{A})^{-1}$. Note that the $n_{chan}\times n_{chan}$ matrix $\boldsymbol{N}$ here (to be distinguished from $\boldsymbol{N_\ell}$) is the noise covariance \emph{at each pixel}, whose diagonal element is, $N_{\nu\nu}=\tfrac{(12\times nside^2)}{4\pi}\times\big(\Delta\Omega\sigma_\nu^2\big)$.
For three known component template maps (i.e., $s_{cmb}$, $s_{sync}$ and $s_{dust}$), the $n_{comp}\times n_{comp}$ matrix $\boldsymbol{\hat{F}}$ in Eq. \eqref{eq-Sigma-for-beta} is, 
\begin{equation}\label{eq-F-hat-matrix}
\big(\boldsymbol{\hat{F}}\big)_{ij}=\sum_p\boldsymbol{s}_i^p\boldsymbol{s}_j^p ~,
\end{equation}
where $i,j=cmb$, $sync$ or $dust$, and the superscript $p$ denotes the pixel location.

To calculate the matrix $\boldsymbol{\Sigma}$, we need to have the synchrotron and the dust polarization template maps (i.e., $s_{sync}$ and $s_{dust}$), and a mask that specifies $nside$ and which pixels are included in the sum in Eq. \eqref{eq-F-hat-matrix}. We do not actually need a template map for the CMB. That is because $A_{cmb}^{raw}$ does not depend on $\beta_s$ or $\beta_d$, and the corresponding CMB components do not contribute to the summation when we take the trace in eq \eqref{eq-F-hat-matrix}. In this work, we use the second Planck release of component polarization maps and the polarization mask, and we degrade them to $nside=128$ resolution. 
Once the matrix $\boldsymbol{\Sigma}$ is obtained, the foreground residuals can be computed as,
\begin{equation}\label{Sigma_matrix}
C_\ell^{fg,res}=\sum_{\beta\beta'}\sum_{jj'}\Sigma_{\beta\beta'}\kappa_{\beta\beta'}^{jj'}C_\ell^{jj'}~,
\end{equation}
where $\kappa_{\beta\beta'}^{jj'}$ is given by,
\begin{equation}
\kappa_{\beta\beta'}^{jj'}=a_\beta^{0j}a_{\beta'}^{0j'}~,
\end{equation}
and $a_\beta^{0j}$ is,
\begin{equation}
a_\beta^{0j}=\left[\boldsymbol{C}_N\boldsymbol{A}^T(\boldsymbol{N})^{-1}\frac{\partial \boldsymbol{A}}{\partial \beta}\right]^{0j}~.
\end{equation}
The $C_\ell^{jj'}$'s in Eq. \eqref{Sigma_matrix} are the auto and cross power spectra of the synchrotron and dust polarization maps.

\begin{figure}[!tp]
\includegraphics[width=0.48\textwidth,height=0.3\textwidth]{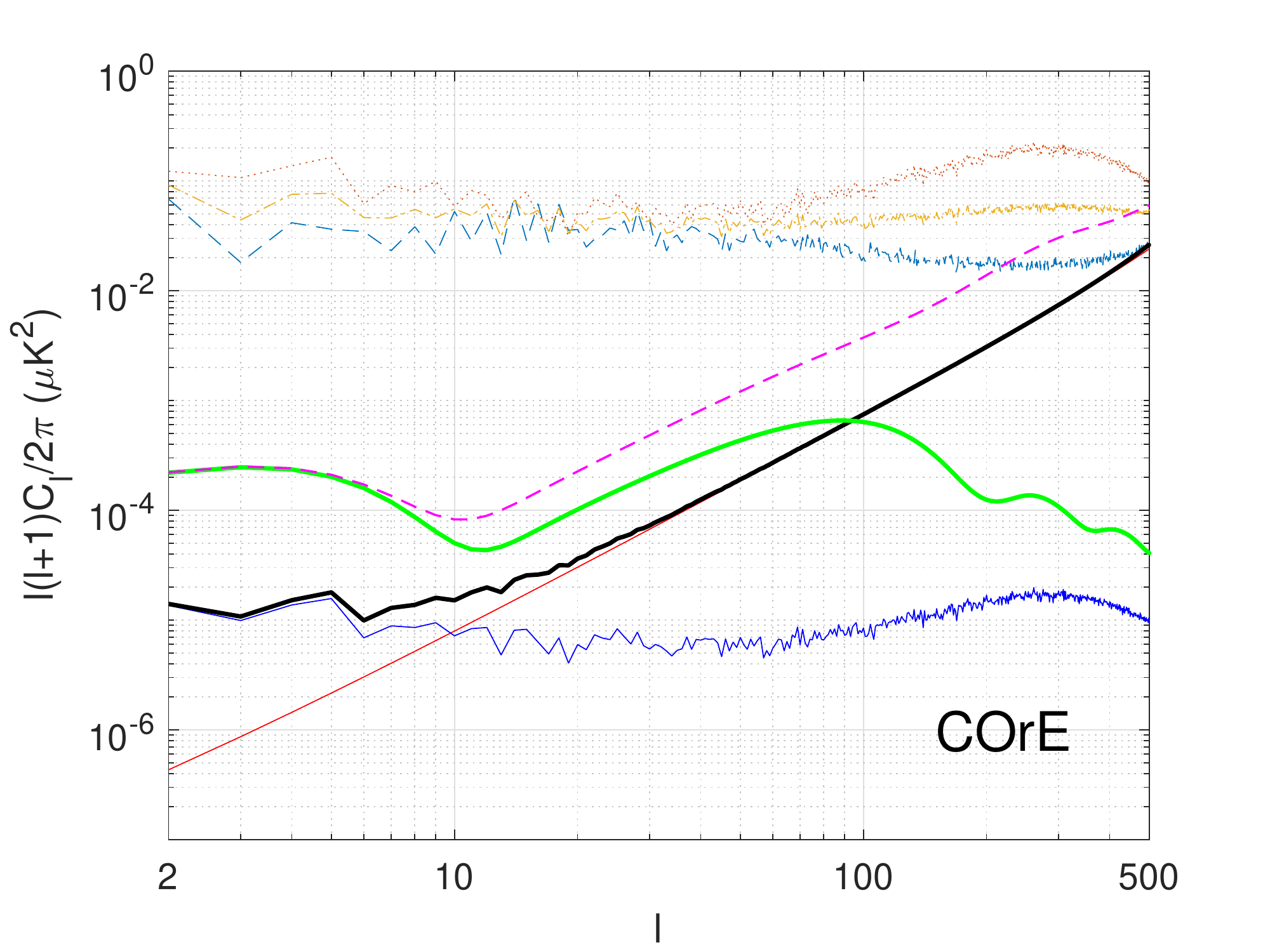}
\includegraphics[width=0.48\textwidth,height=0.3\textwidth]{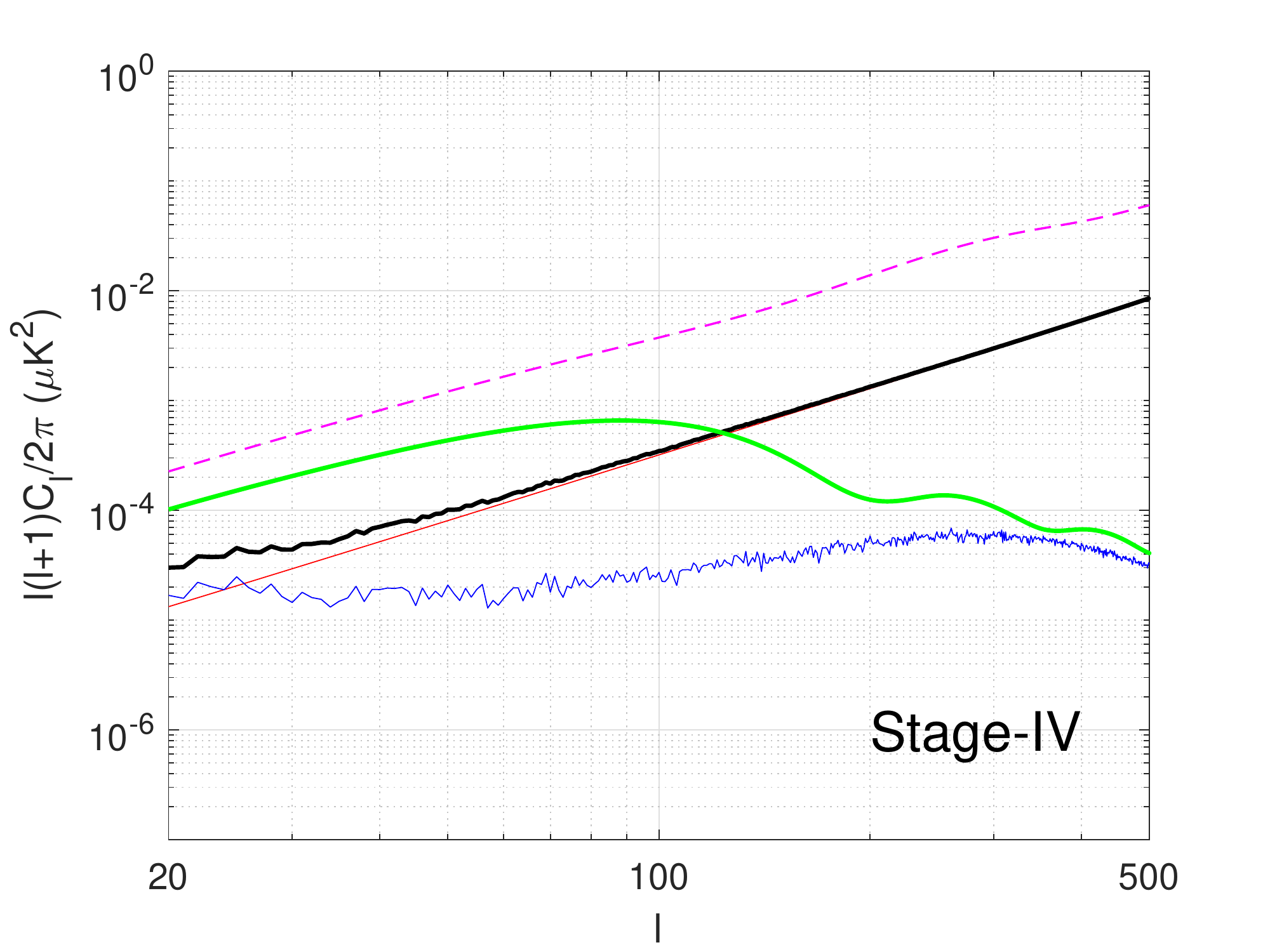}
\includegraphics[width=0.48\textwidth,height=0.3\textwidth]{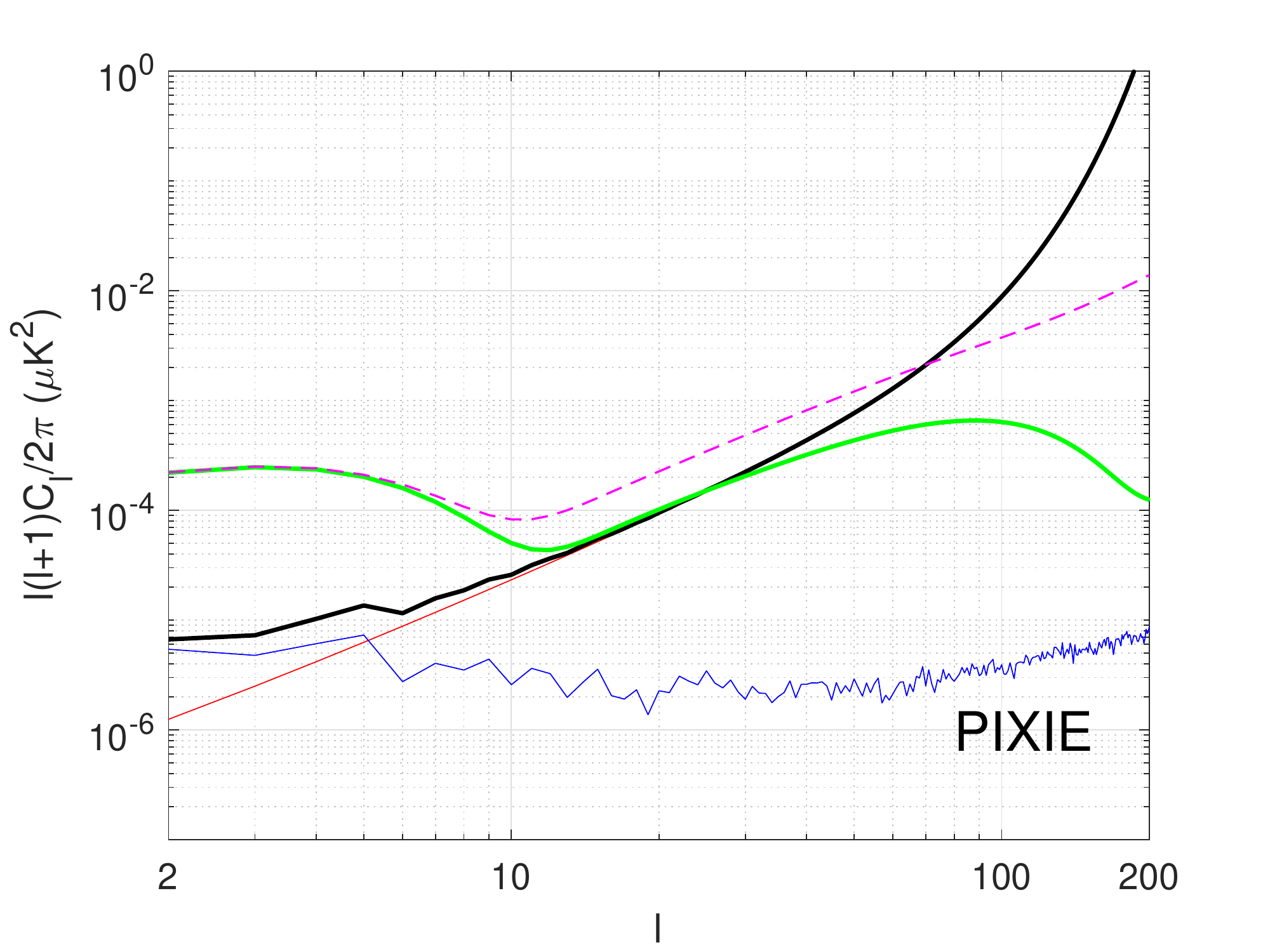}
\caption{\label{fig-Cls}COrE (top), Stage-IV (middle) and PIXIE (bottom): The power spectra of 1) the tensor B-mode polarization with $r=0.01$ in $\Lambda$CDM (solid green), 2) the total B-mode (dash magenta), 3) the degraded instrumental noise (solid red), 4) the (total) foreground residual (solid blue), 5) the total noise-like error (solid black), and 6) the foreground signals (shown only on the top of the COrE panel: dotted for synchrotron-auto, dashed for dust-auto and dot-dashed for synchrotron-dust cross spectra). Note the minimal $\ell$ for Stage-IV is just $20$. And the maximum $\ell$ for PIXIE is $200$.}
\end{figure}

We refer readers to Refs. \cite{B-mode-forecast-Errard2011,CMB-Bmode-forecast2015} for detailed discussions of the above framework. In Fig. \ref{fig-Cls} we show results for the power spectra of the degraded instrumental noise, the (total) foreground residual and the B-mode polarization with our base fiducial model for the three future experiments we considered. Different experiment specifications lead to different degraded noises and foreground residuals.

\subsection{Performance forecast of constraints on tensor-mode MG parameters}\label{subsection-forecast-relsults}
In this subsection, we consider the following question: how significant do the deviations from GR in the tensor sector need to be, so that we can detect them with the near-future CMB experiments? To answer this question, we do a performance forecast using the Fisher matrix formalism with the specifications of COrE, Stage-IV and PIXIE listed in Tables \ref{COrEspec}, \ref{S4spec} and \ref{PIXIEspec}.

\begin{table*}[t]
\begin{ruledtabular}
\caption{\label{fiducial_model}The base fiducial model ($\Lambda$CDM $+r$) used in the Fisher matrix analysis. We extend it to four MG models (i.e. $\Lambda$CDM $+\,r\,+$ 1 MG parameter).}
\begin{tabular}{lccccccc}
Base fiducial parameters & $r$ & $n_s$ & $\tau$ & $\Omega_bh^2$  &$\Omega_ch^2$  & $H_0$ & $A_s$\\
\hline
Values & $0.01$ &$0.9645$ &$0.079$& $0.02225$ & $0.1198$ & $67.27$ & $2.2065\times10^{-9}$\\
\end{tabular}
\end{ruledtabular}
\end{table*}

\begin{figure*}[tbp]
\centering
\includegraphics[width=0.325\textwidth,height=0.3\textwidth]{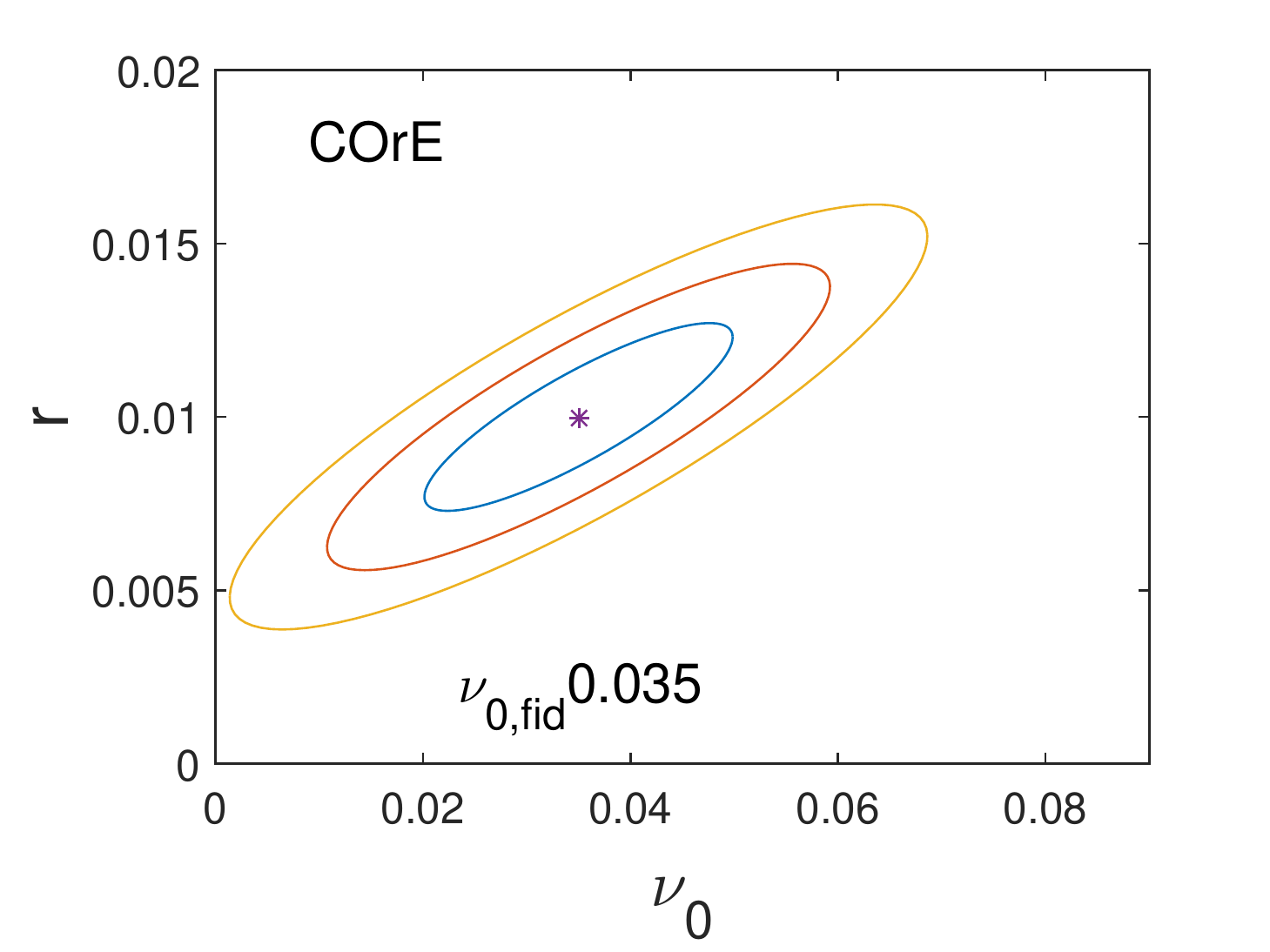}
\includegraphics[width=0.325\textwidth,height=0.3\textwidth]{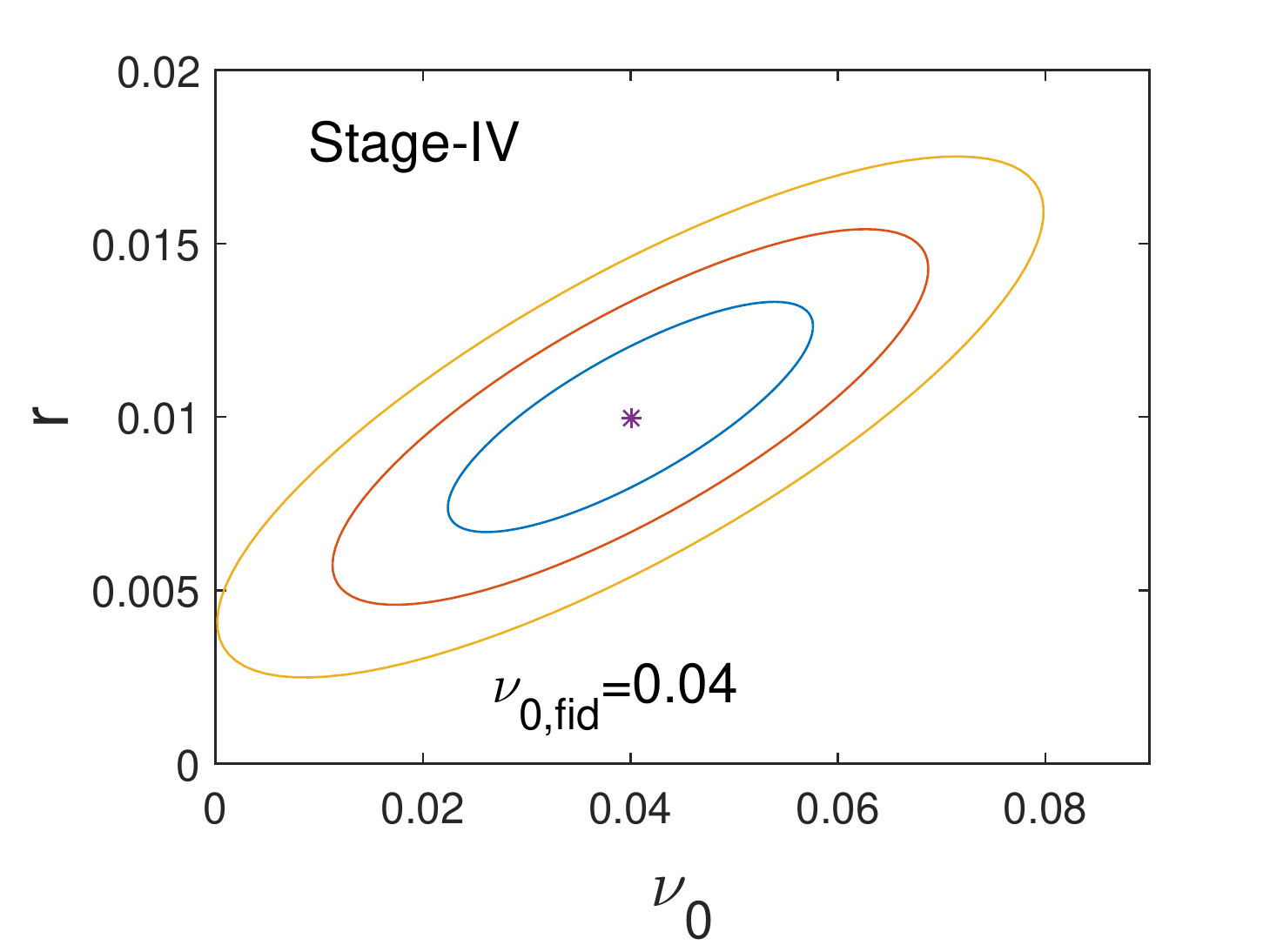}
\includegraphics[width=0.325\textwidth,height=0.3\textwidth]{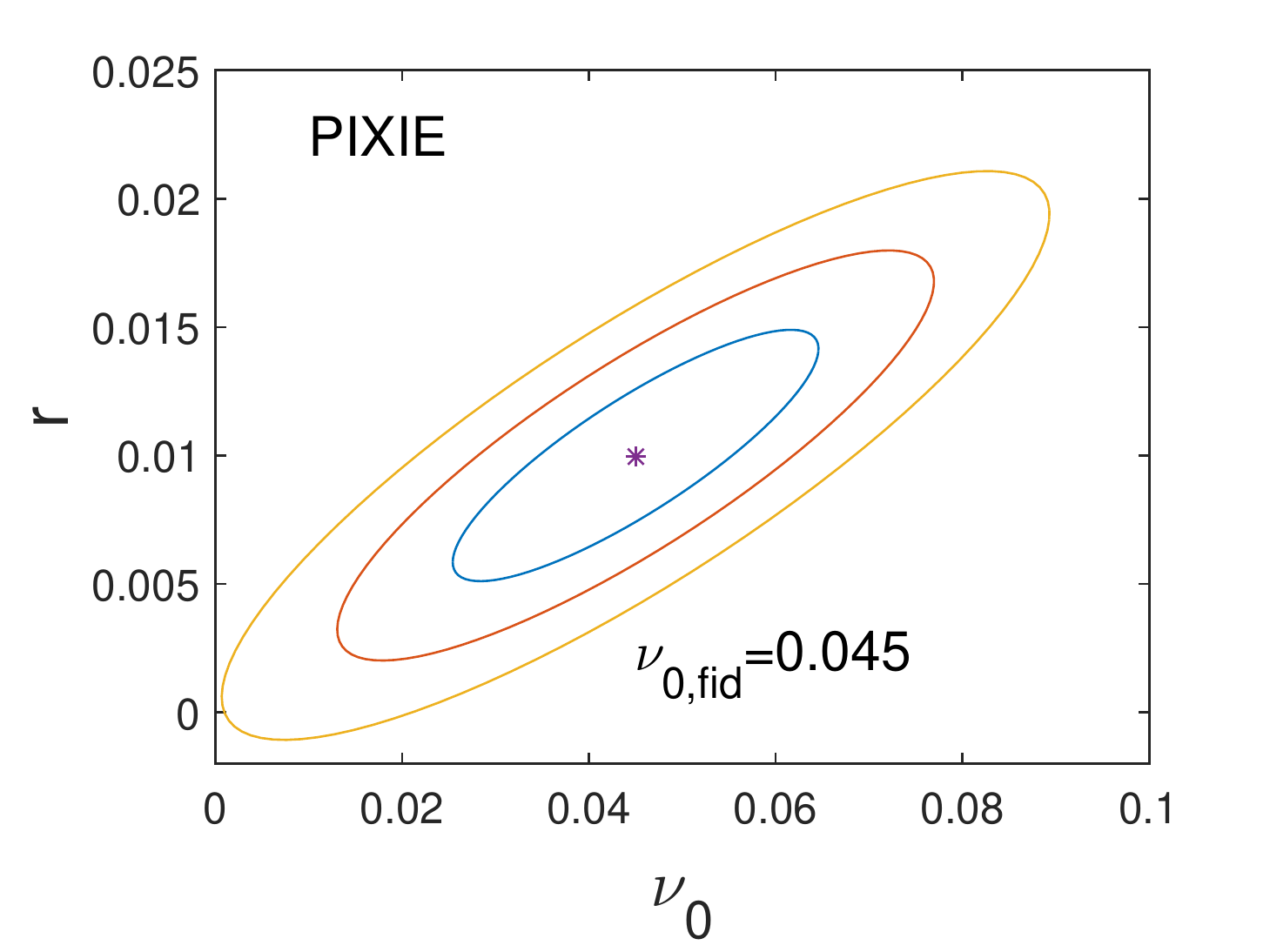}
\includegraphics[width=0.325\textwidth,height=0.3\textwidth]{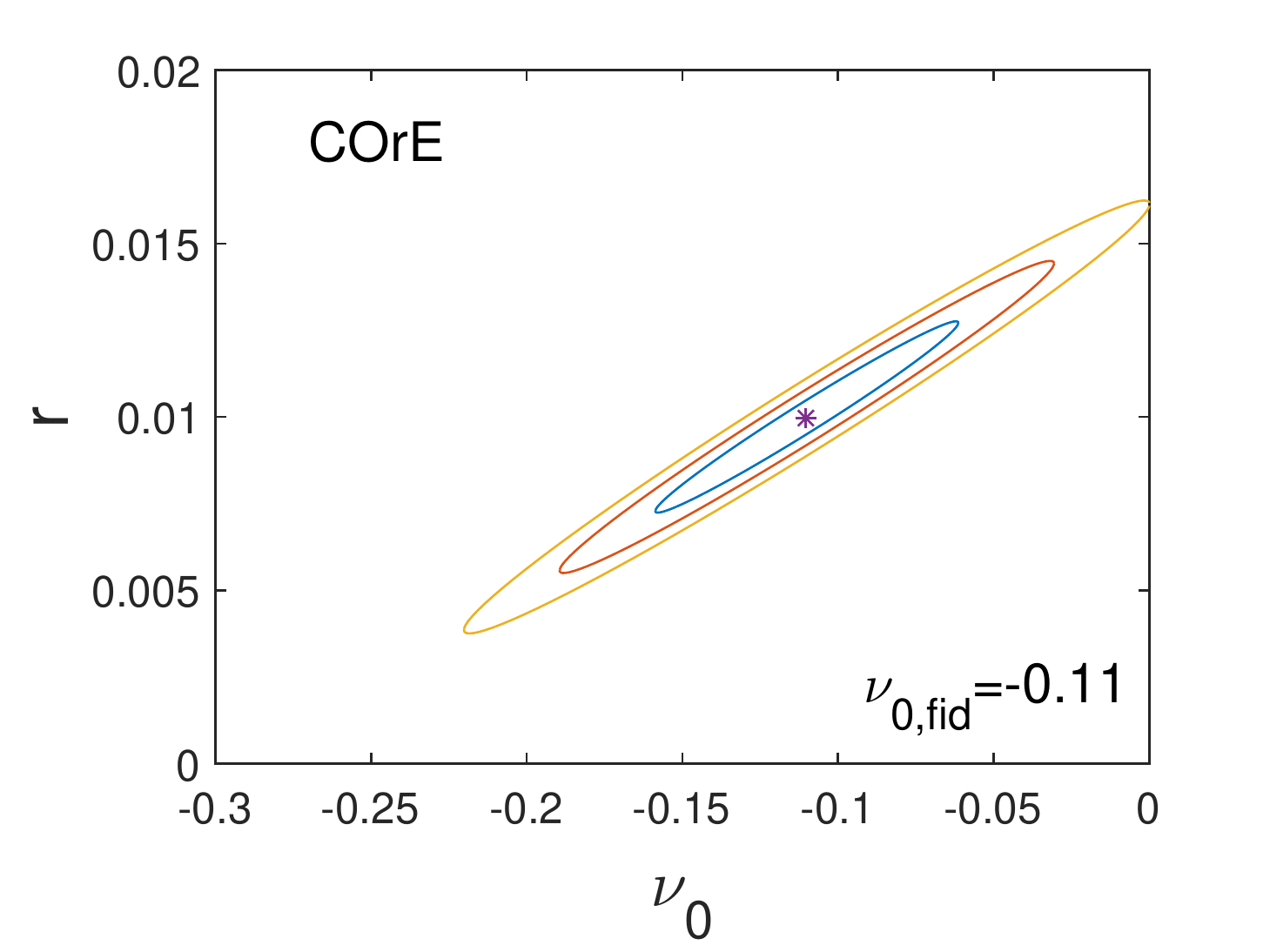}
\includegraphics[width=0.325\textwidth,height=0.3\textwidth]{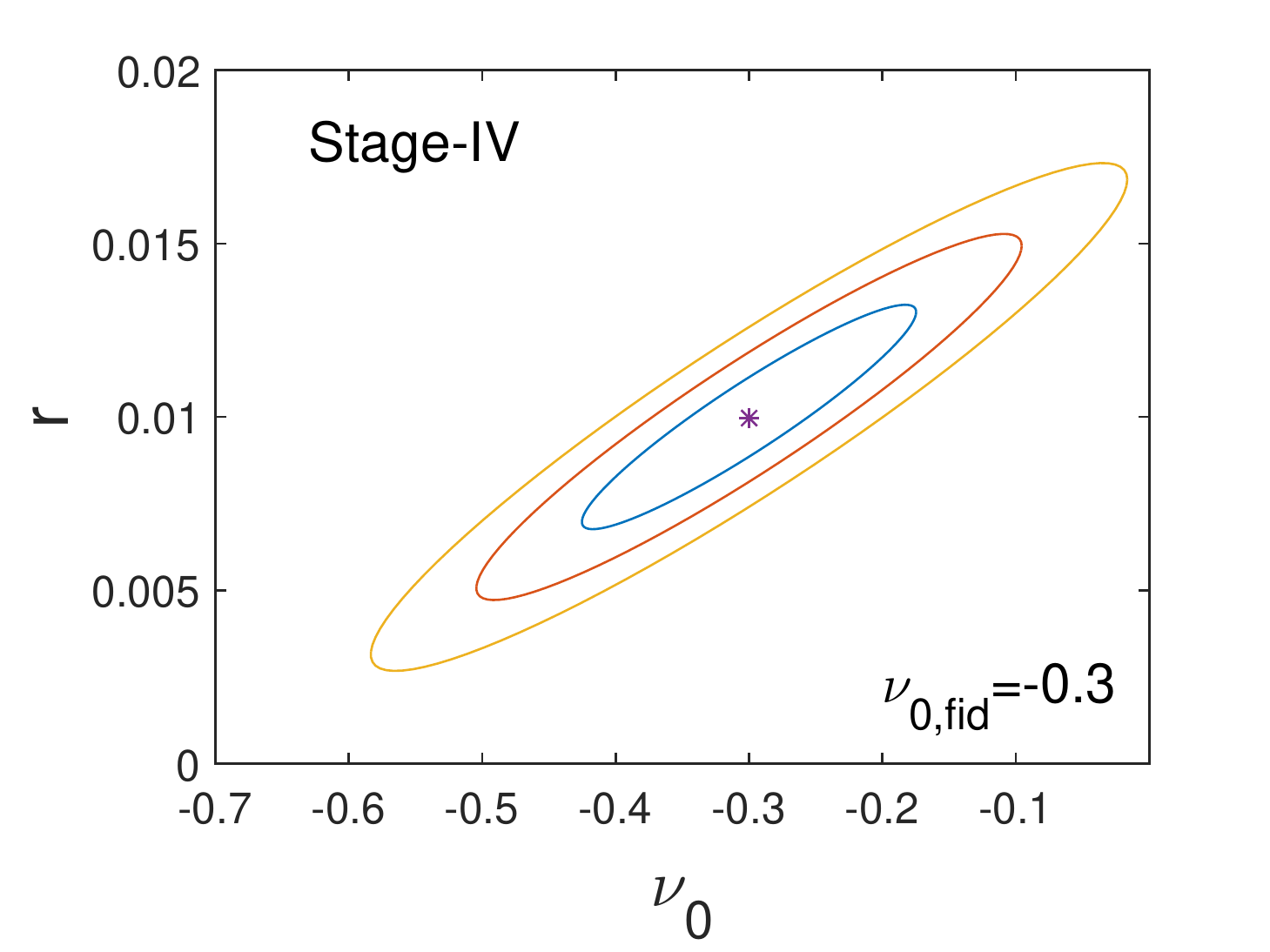}
\includegraphics[width=0.325\textwidth,height=0.3\textwidth]{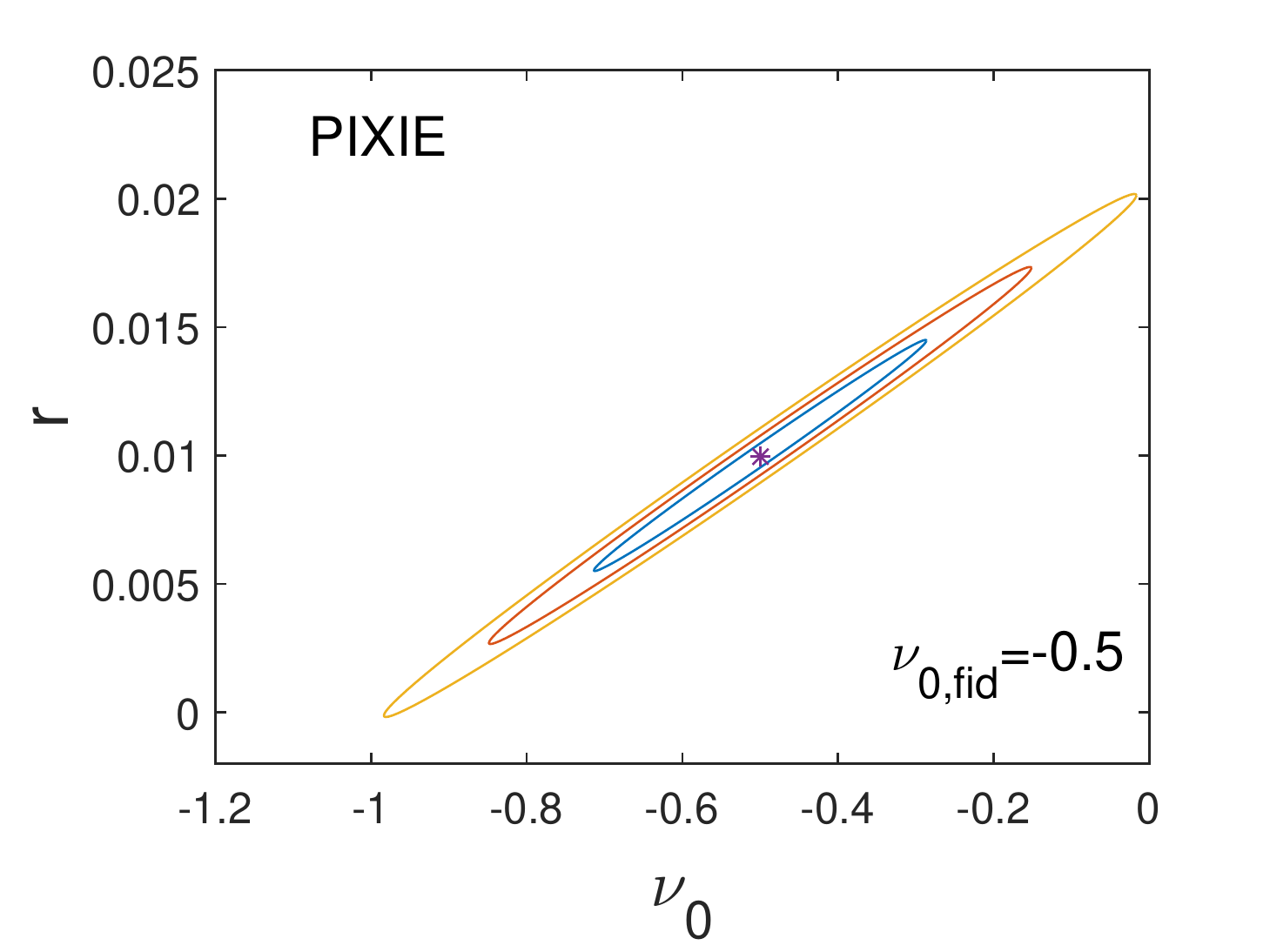}
\caption{\label{r_nu0_min}Results of constraints on the friction term for COrE (left), Stage-IV (middle) and PIXIE (right) specifications. We show the $1$-$\sigma$, $2$-$\sigma$ and $3$-$\sigma$ marginalized confidence-region contours in the $r$-$\nu_0$ space for the $\Lambda$CDM+$r$+$\nu_0$ model. We set $r_{fid}=0.01$. All top (bottom) panels are for positive (negative) $\nu_0$. These figures show the minimum detectable values of $\nu_0$, which can be converted to a minimally required percentage difference in the strength of friction.}
\end{figure*}

In Table \ref{fiducial_model} we list the base fiducial model used in our Fisher matrix analysis. In this subsection, we only consider the $\Lambda$CDM+$r$ with the standard inflation consistency relation as our base model, where $\Lambda$CDM stands for the six standard cosmological parameters. The test of the standard vs the MG consistency relation will be in the next subsection. On top of the base model, we consider four extended models, namely, $\Lambda$CDM+$r$+$\nu_0$, $\Lambda$CDM+$r$+$\varepsilon_0$, $\Lambda$CDM+$r$+$\varepsilon_l$, and $\Lambda$CDM+$r$+$\varepsilon_h$. When we consider the $\Lambda$CDM+$r$+$\nu_0$ model, for example, we fix the other MG parameters to their GR values. The six standard $\Lambda$CDM parameters are then marginalized over to give two-dimensional confidence-region plots in the $r$ + $\nu_0$. We then derive the minimum detectable values of the tensor-mode MG parameters for those future experiments. In this work, the minimum detectable value $x_{min}$ of an MG parameter $x$ is conservatively defined as the one when the $x$-direction half width of the $3$-$\sigma$ likelihood ellipse in the marginalized $r$-$x$ space equals $x_{min}$ itself (or $-x_{min}$ if $x$ is negative). We will repeat and do the same for the other extended models. These minimum detectable values should depend on the base fiducial model, especially on the fiducial value of $r$. We do not consider the constraints on MG parameters simultaneously since the near-future CMB experiments all have limited constraining power. Moreover, we want to explore the individual minimum detectable value for each MG parameter so we can estimate which modification to GR will be most likely detectable with these experiments.

\begin{figure*}[htbp!]
\begin{center}
\includegraphics[width=\textwidth,height=0.72\textheight]{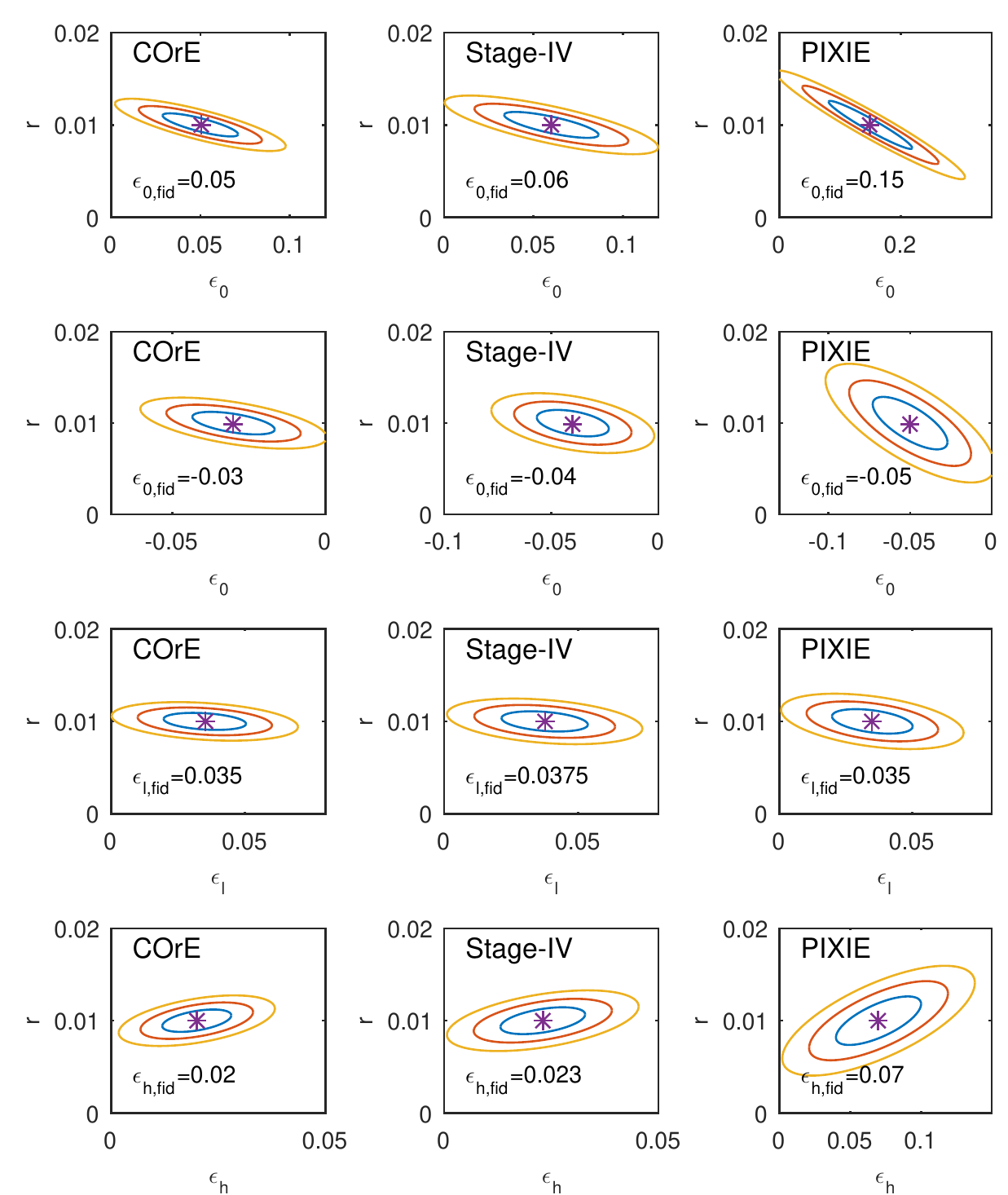}
\caption{\label{r-epsilons}Results of constraints on the dispersion relation for the COrE (left), Stage-IV (middle) and PIXIE (right) specifications. First two rows: the $\Lambda$CDM+$r$+$\varepsilon_0$ model. Take COrE for example: a value of $|\varepsilon_{0,min}|=0.05$ means COrE can observe a speed fractional deviation that is $5\%$ different from the speed of light. Third row: the $\Lambda$CDM+$r$+$\varepsilon_l$ model. A value of $\varepsilon_{l,min}=0.035$ (with $n=4$) means the minimum detectable mass of the graviton will (at best) be $7.8\times10^{-33}\,eV$. Fourth row: the $\Lambda$CDM+$r$+$\varepsilon_h$ model. This is a high-$k/a$ deviation model, $\varepsilon_{h,min}=0.02$ means the dispersion is not changed for a physical wave number smaller than $k_0/\sqrt{\varepsilon_h}=700$ Mpc$^{-1}$. Similar interpretations apply to the other two experiments.}
\end{center}
\end{figure*}

\begin{table*}[!htbp]
\begin{ruledtabular}
\caption{\label{table_result_list}Results for the COrE specifications of the minimum detectable values of the tensor mode modified gravity parameters and their physical meaning with $r=0.01$.}
\begin{tabular}{ldp{0.7\textwidth}}
  $\Lambda$CDM$+r+$ & \multicolumn{1}{R{0.1\textwidth}}{Minimum detectable} & Physical effects associated with a  detection at the $3$-$\sigma$ level\\
\hline
   $\nu_0$ & 0.035 & An enhanced friction that is $3.5\%$ (or more) larger than that in GR can be detected\\
negative $\nu_0$ & -0.11 & A suppressed friction that is at least $11\%$ smaller than the GR value can be detected\\
$|\varepsilon_0|$ & 0.04 & A speed deviation from the speed of light of $\sim4\%$ or larger can be detected\\
$\varepsilon_l$ & 0.035 & A graviton mass $>7.8\times10^{-33}\,eV$ can be detected\\
$\varepsilon_h$ & 0.02 & The small-scale dispersion relation needs to be modified with a critical wave number $ (k/a)_{critical} \lesssim 700$ Mpc$^{-1}$ (or critical scale $\gtrsim1.4$ kpc) for detection.\\
\end{tabular}

\caption{\label{table_result_list_PIXIE}Results for the Stage-IV specifications, similar to Table\ref{table_result_list}.}
\begin{tabular}{ldp{0.7\textwidth}}
  $\Lambda$CDM$+r+$ & \multicolumn{1}{R{0.1\textwidth}}{Minimum detectable} & Physical effects associated with a  detection at the $3$-$\sigma$ level\\
\hline
$\nu_0$ & 0.04 & An enhanced friction that is $4\%$ (or more) larger than that in GR can be detected\\
negative $\nu_0$ & -0.3 & A suppressed friction that is at least $30\%$ smaller than the GR value can be detected\\
$|\varepsilon_0|$ & \sim0.05 & A speed deviation from the speed of light of $\sim5\%$ or larger can be detected\\
$\varepsilon_l$ & 0.038 & A graviton mass $>9.7\times10^{-33}\,eV$ can be detected\\
$\varepsilon_h$ & 0.023 & The small-scale dispersion relation needs to be modified with a critical wave number $ (k/a)_{critical}\lesssim 660$ Mpc$^{-1}$ (or critical scale $\gtrsim1.5$ kpc) for detection.\\
\end{tabular}

\caption{\label{result_list_PIXIE}Results for the PIXIE specifications, similar to Table\ref{table_result_list}.}
\begin{tabular}{ldp{0.7\textwidth}}
$\Lambda$CDM$+r+$ & \multicolumn{1}{R{0.1\textwidth}}{Minimum detectable} & Physical effects associated with a  detection at the $3$-$\sigma$ level\\
\hline
 $\nu_0$ & 0.045 & An enhanced friction that is $4.5\%$ (or more) larger than that in GR can be detected\\
negative $\nu_0$ & -0.5 & A suppressed friction that is at least $50\%$ smaller than the GR value can be detected\\
$|\varepsilon_0|$ & \multicolumn{1}{R{0.1\textwidth}}{$0.15\, \&\, 0.05$} & A speed deviation from the speed of light that is $15\%$ faster, or $5\%$ slower can be detected\\
$\varepsilon_l$ & 0.035 & A graviton mass $>7.8\times10^{-33}\,eV$ can be detected\\
$\varepsilon_h$ & 0.07 & The small-scale dispersion relation needs to be modified with a critical wave number $ (k/a)_{critical}\lesssim 380$ Mpc$^{-1}$ (or critical scale $\gtrsim2.6$ kpc) for detection.\\
\end{tabular}

\end{ruledtabular}
\end{table*}

In Fig. \ref{r_nu0_min} (for friction) and Fig. \ref{r-epsilons} (for dispersion relation) we show the results of the performance forecast. Take the COrE specification for example: we can infer from those plots that the minimum detectable values of $\nu_0$, $|\varepsilon_0|$, $\varepsilon_l$, and $\varepsilon_h$ are $0.035$ ($-0.11$ for negative $\nu_0$), $\sim0.05$, $0.035$ and $0.02$ respectively. These minimum detectable values tell us that the COrE mission can detect deviations from GR if 1) the additional friction is at least $3.5\%$ larger than that in GR, 2) or the friction is suppressed and at least $11\%$ less than that in GR, 3) the speed of gravitational waves is at least by $\sim5\%$ different from the speed of light, 4) gravitons possess a mass of at least $7.8\times10^{-33}\,eV$, and 5) the small-scale dispersion relation is modified with a critical scale of $1.4$ kpc. The critical scale in the last case is defined as the inverse of $k_0/\sqrt{\varepsilon_h}$, which means the dispersion relation at scales smaller than this will be modified. In particular, the $\Lambda$CDM+$r$+$\varepsilon_l$ model corresponds to a massive graviton model. With $r=0.01$ and the standard inflation consistency relation, the minimum detectable graviton mass is $7.4\times10^{-33}\,eV$ for COrE. This is important, because, as we mentioned earlier, if the massive gravity models are responsible for the late-time cosmic acceleration, the graviton mass will be at the order of $10^{-33}\,eV$.

The minimum detectable graviton mass depends on the value of $n$ we set in Eq. \eqref{eq-dispersion-cases}. We set $n=4$ for convenience in the MCMC analysis with the current data. We can choose a different $n$ for future data. Choosing a different $n$ will give us a different value of $\varepsilon_{l,min}$, and consequently a different minimum detectable graviton mass. This is because changing the value of $n$ effectively sets a different uniform prior. But this change does not give a very different result. For example, we later set $n=1$ and obtain a minimum detectable graviton mass of $8.5\times10^{-33}\,eV$.

We list all the minimum detectable values and their physical meanings in Table \ref{table_result_list} for the three near-future experiments. We found that those three near-future experiments are optimistic about the constraints of the tensor-mode MG parameters. For $r_{fid}=0.01$, the additional friction only needs to be different from that in GR by $3.5\sim4.5\%$ to allow detection. If the friction is suppressed (negative $\nu_0$), it is required to be $11\sim50\%$ smaller than that in GR for detection. For the speed of gravitational waves, it only requires a difference of $4\sim15\%$. All experiments can detect a graviton mass with a magnitude of the order of $10^{-33}\,eV$, comparable to the one in the massive gravity theories that give late-time cosmic acceleration.

At the end of this subsection, it is worth clarifying why we can constrain $\varepsilon_h$ in the presence of lensing. It is true that $\varepsilon_h$ only changes the tensor-induced B-mode power spectrum at small scales, where it is generally considered to be contaminated by the signal from lensing. But if the tensor-to-scalar ratio $r$ is not completely negligible, the tensor-mode contributions are important for B-mode polarization at $\ell\lesssim150$. A larger $\varepsilon_h$ leads to a smaller $\ell$ onset of the damping effects on the B-mode power spectrum; see Fig. \ref{fig-high-camb}. The values of the minimum detectable $\varepsilon_h$ shown in Tables \ref{table_result_list}, \ref{table_result_list_PIXIE} and \ref{result_list_PIXIE} are large compared to the ones shown in Fig. \ref{fig-high-camb}, which are large enough to suppress the B-mode power spectrum within $\ell\lesssim150$. If the foreground signals can be truly subtracted down to the levels shown in Fig. \ref{fig-Cls}, we will be able to see this suppressing effect due to the MG parameter $\varepsilon_h$.

\begin{figure*}
\includegraphics[width=0.49\textwidth]{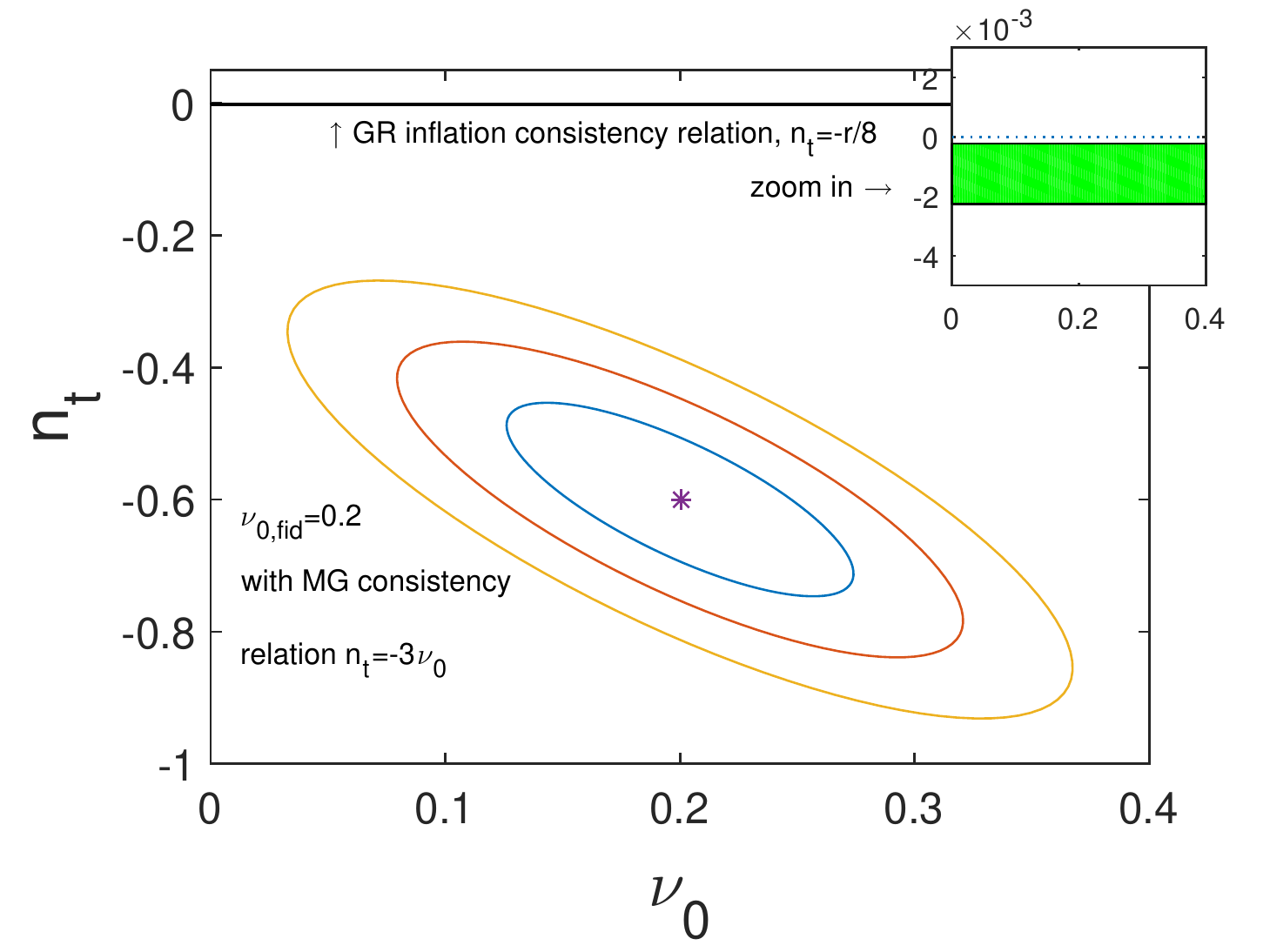}
\includegraphics[width=0.49\textwidth]{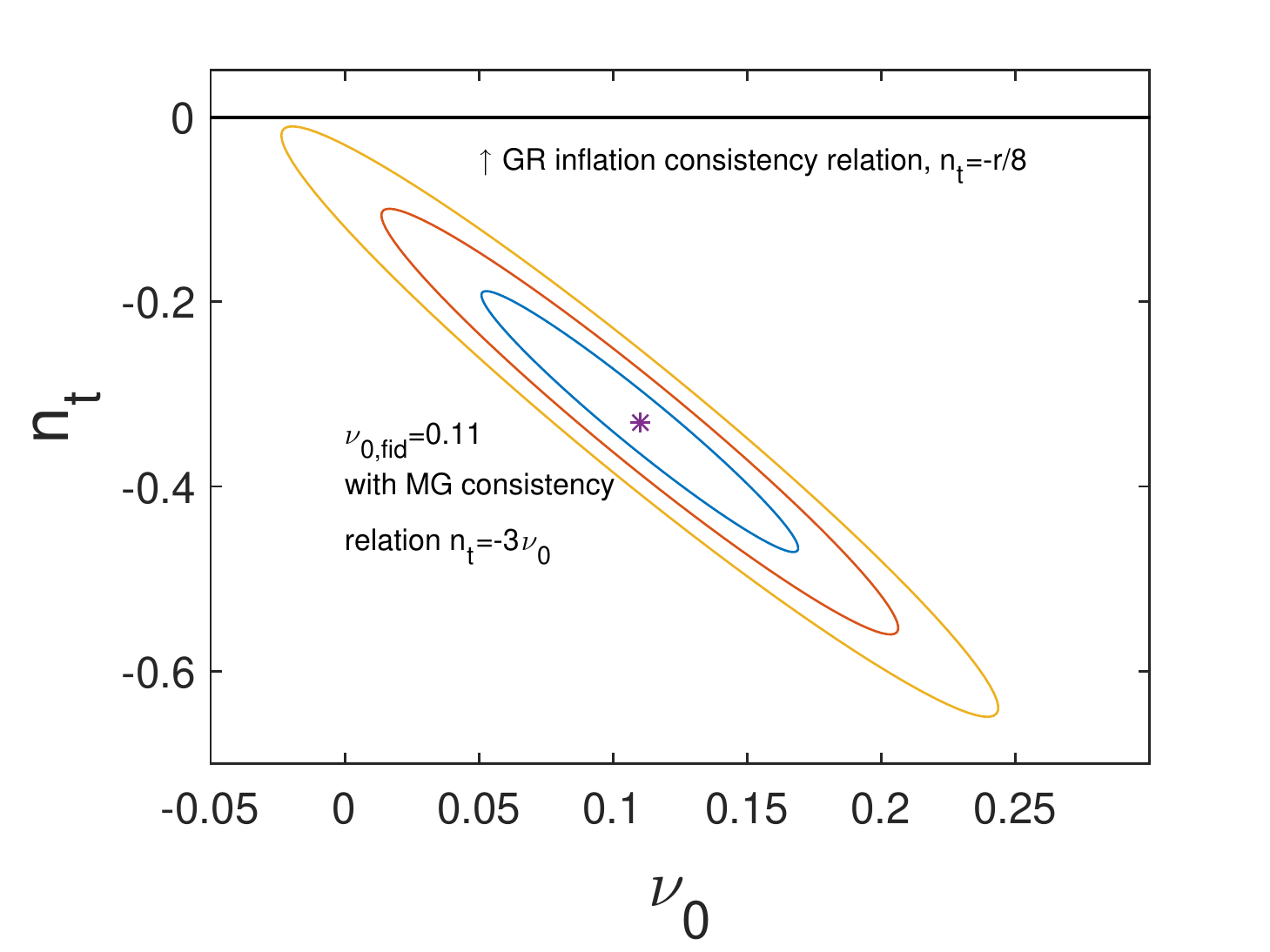}
\caption{\label{MG_consistency_relation}Demonstration of how we can distinguish the standard and the MG consistency relations. We assume that the fiducial model satisfies the MG consistency relation with $\nu_0=0.2$ on the left and $\nu_0=0.11$ on the right. Both panels have a fiducial value of $r=0.01$. For the left panel, the MG consistency relation predicts $n_T\simeq-0.6$, which is much larger than the one predicted in GR ($n_T=-0.00125$) with the standard consistency relation. There is a shaped band in the figure that shows the range of $n_T$ according to the standard consistency relation $n_T=-r/8$. That shaped band is so narrow that it looks like a ``straight line'' in the $\nu_0$ vs $n_T$ parameter space. The side box shows the shaped band with a $3$-$\sigma$ uncertainty of $r$ in a more suitable range. We can see that the three iso-likelihood contours \emph{do not} intersect with the shaped band. Therefore, such simulated data favor the MG consistency relation over the standard consistency relation. However, the true value of $\nu_0$ needs to be large enough in order to distinguish the two consistency relations observationally. The right panel shows the minimum value of $\nu_0$ that allows us to distinguish the two consistency relations for COrE, which is $\nu_{0,min}=0.11$.}
\end{figure*}

\subsection{Testing the standard consistency relation vs the MG consistency relation}\label{subsection-forecast-cons-rel}
Another question is: can we test the standard consistency relation \eqref{eq-slow-roll-consistency-rel} vs the MG consistency relation \eqref{eq-MG-consistency-rel}?  We find that in some situations we are able to do so, and we show it with the method of performance forecast described in the previous subsection. We assume in this work that the friction parameter $\nu_0$ is constant throughout the history of the Universe.

We first extend the model from $\Lambda$CDM+$r$+$\nu_0$ to  $\Lambda$CDM+$r$+$\nu_0$+$n_T$, where $n_T$ is the tensor spectral index. We assume the true value of $\nu_0$ is much larger than $r$. Here we set $r_{fid}=0.01$ and $\nu_{0,fid}=0.2$. The small term $-r/8$ can be ignored in the MG consistency relation \eqref{eq-MG-consistency-rel}, so it becomes $n_T\simeq-3\nu_0=-0.6$. On the other hand, the standard consistency relation gives $n_T=-r/8=-0.000125$. Therefore, the two consistency relations can be very different: while $|n_T|$ can be large for the MG consistency relation, it must be small for the standard one (given the fact that $r<0.1$ from current observational upper bound). To experimentally test the two consistency relations, we want to see whether future data are consistent with only one of them. In our performance forecast, we set the fiducial model to be consistent with the MG consistency relation. At the end, we will marginalize over the six standard $\Lambda$CDM parameters and $r$ to get a two-dimensional confidence-region plot in the $n_T$ vs $\nu_0$ parameter space. Once we obtain such a two-dimensional plot, we will be able to see whether the uncertainty is small enough to rule out the standard consistency relation.

We take the COrE as an example to examine the above question. In the left panel of Fig. \ref{MG_consistency_relation}, the co-center of  the three ellipses shows the fiducial model in the $n_T$ vs $\nu_0$ parameter space, and the three ellipses are the $1$-$\sigma$, $2$-$\sigma$ and $3$-$\sigma$ marginalized likelihood contours. The ``straight line'' shows the standard consistency relation $n_T=-r/8$ with $3$-$\sigma$ uncertainty of $r$. This ``straight line'' is actually a green shaped band. But its offset from $0$ and its uncertainty are too small compared to the vertical scale of the graph, so it looks like a straight line. We zoom in and show this shaped band in a side box in the top-right corner.  The ellipses do not intersect with the shaped band, which means the observation is not consistent with the standard consistency relation at the $3$-$\sigma$ confidence level. In such a case, we can verify the MG consistency relation and rule out the standard one.

The next question is: how large does $\nu_0$ need to be for us to experimentally distinguish the two consistency relations? If the fiducial value of $\nu_0$ is small, $n_T$ will also be small even if it follows the MG consistency relation. The ellipses will then move upwards in the $r$ vs $\nu_0$ plane, and intersect with the shaped band. In that case the data will be consistent with both consistency relations, and we will not be able the tell which one is correct. The minimum value of $\nu_0$ (for COrE) that allows us to observationally distinguish the two consistency relations (at the $3$-$\sigma$ C.L.) is demonstrated in the right panel of Fig. \ref{MG_consistency_relation}. There we set the fiducial value of $\nu_0$ to $0.11$. The $3$-$\sigma$ likelihood contour marginally intersects with the shaped band. So if $\nu_0>0.11$, the ellipses will be below the shaped band (like the case in the left panel), and if $\nu_0<0.11$ they intersect. This minimum value of $\nu_0$ is still very large compared to $r$, that is, $\nu_{0,min}=0.11\gg r=0.01$.

For the case of negative $\nu_0$, the discussion will be similar to that above. But since the negative $\nu_0$ is more difficult to observe (see Sec. \ref{subsection-forecast-relsults}), $|\nu_0|$ needs to be very large for us to distinguish the standard and the MG consistency relations.

The conclusion of this subsection is that: yes, in some situations, we can observationally distinguish the standard and the MG consistency relations. The friction parameter $|\nu_0|$ needs to be much larger than the tensor-scalar-ratio $r$ in order for us to experimentally disentangle the standard and the MG consistency relations with the next-generation CMB experiments.

\section{Summary}\label{section-summary}
We proposed a general form of the tensor-mode propagation equation, which can be applied to a wide range of modified gravity theories. Based on this equation, we wrote four physically motivated parametrization schemes which include the changes to the friction, the propagation speed, as well as the dispersion relation at large and small scales. Some similar modifications have been individually considered in the literature \cite{friction,MG.Ct.of.GW,MassiveGinCMB}, but we combined them in a different approach and extend them to cover more possible cases. We also derived a consistency relation for the MG models. We then performed parameter constraints and forecasts.

Before investigating the current and future data constraints, we studied the parametrized tensor-mode perturbations during inflation and derived a few useful equations in the modified gravity case. We obtained an MG inflation consistency relation $n_T=-3\nu_0-r/8$. Besides relating the tensor spectral index $n_T$ to the tensor-to-scalar ratio $r$ as in the standard inflation consistency relation, the MG inflation consistency relation also relates $n_T$ to the friction parameter $\nu_0$. If the friction parameter is constant throughout the history of the Universe (including inflation and the period after it), we can use the CMB B-mode polarization data to test the standard and the MG consistency relations. If the friction parameter is finite but changes its value after inflation, then at least the standard inflation consistency relation can be falsified due to the additional contribution from $\nu_0$ to the value of $n_T$.

To see the MG effects on the B-mode polarization and to constrain the MG parameters from the current observations, we modify \textsc{camb} to implement our parametrization and apply a Monte Carlo Markov Chain analysis using \textsc{CosmoMC}. We studied the effects of the four parameters individually on the B-mode polarization power spectrum. Then using the currently available data from the Planck-BICEP2 joint analysis and the Planck-2nd-released low-$\ell$ polarization, we set exclusion regions on the MG parameters.

Then we calculated performance forecasts on constraining MG parameters for the next-generation CMB experiments. We used the specifications of the near-future missions COrE, Stage-IV and PIXIE. We performed calculations of the corresponding foreground residuals and the degraded noise for the analysis. For a fiducial cosmological model with a tensor-to-scalar ratio $r=0.01$, we determined the 3-$\sigma$ confidence contours in the $r~+$ each MG parameter spaces. We found that (\textit{i}) an additional relative friction of $3.5\sim4.5\%$ compared to its GR value will be detected at the $3$-$\sigma$ level by these experiments (the details are given in our Tables \ref{table_result_list}, \ref{table_result_list_PIXIE}, and \ref{result_list_PIXIE}); (\textit{ii}) a suppressed friction will be harder to constrain ($-11$ to $-50\%$ is required for a detection); (\textit{iii}) the speed of gravitational waves with a relative difference of $5\sim15\%$ or larger compared to the speed of light will be detected; (\textit{iv}) the minimum detectable graviton mass is around $7.8\sim9.2\times10^{-33}\,eV$ for these experiments: this is important because this minimum detectable graviton mass is of order of $10^{-33}\,eV$, which is the same as the one in the massive gravity theories that can produce the late-time cosmic acceleration; (\textit{v}) for the small-scale deviation, the dispersion relation needs to be modified with a critical wave number $( k/a)_{critical}\lesssim 380\sim700$ Mpc$^{-1}$ (or the critical scale needs to be $\gtrsim 1.4\sim2.6$ kpc) for detection.

Finally, with the performance forecast, we explored the possibility for the next-generation CMB experiments to distinguish the MG inflation consistency relationship ($n_T=-3\nu_0-r/8$) from the standard inflation consistency relation ($n_T=-r/8$). We showed that in order to disentangle the two consistency relations, the MG friction parameter $|\nu_0|$ needs to be much larger than the tensor-to-scalar ratio $r$.

In summary, we find that the near-future experiments probing tensor-induced B-modes such as the COrE mission  \cite{COrEWhitePaper}, PRISM mission \cite{PRISMwhitepaper}, POLARBEAR2 \cite{PolarBear2}, CMB Stage-IV \cite{Stage-IV-paper1} and PIXIE \cite{PIXIE2011}  will open a new promising window on testing gravity theories at cosmological scales.

\begin{acknowledgments}
We would like to thank J. Dossett and E. Linder for useful comments, J. Errard for providing suggestions on the steps to calculate the foreground residuals, H. Eriksen for providing a resolution upgraded version of the Planck synchrotron polarization map, A. Kogut for sending us information on the sensitivity of PIXIE for each channel, and A. Lewis for pointing us to useful references.
MI acknowledges that this material is based upon work supported in part by the NSF under grant AST-1517768 and an award from the John Templeton Foundation.
\end{acknowledgments}


\bibliography{B-mode-BIB}{}

\end{document}